\documentclass[12pt]{article}
\pdfoutput=1
\usepackage[letterpaper, margin=1in]{geometry}
\usepackage{amsmath, amssymb}
\usepackage{cite}
\usepackage{graphicx}
\usepackage[font=small,labelfont=bf]{caption}
\usepackage[normalem]{ulem}
\usepackage{makecell}

\usepackage{slashed}
\usepackage{cancel}
\usepackage{bm}
\usepackage{verbatim}
\usepackage{tabularx}
\allowdisplaybreaks

\usepackage[utf8]{inputenc}
\usepackage{graphicx}
\usepackage{caption}
\usepackage{subfigure}
\usepackage{hhline}
\usepackage{amsmath}
\usepackage{amsfonts}
\usepackage{amssymb}
\usepackage[dvipsnames]{xcolor}
\RequirePackage{luatex85}
\usepackage{braket}
\usepackage{array} 
\usepackage{pict2e}
\usepackage{multirow}
\usepackage[compat=1.1.0]{tikz-feynman}
\usepackage[super]{nth}
\usepackage{float}
\pagecolor{white}
\definecolor{babyblueeyes}{rgb}{0.63, 0.79, 0.95}
\definecolor{carminepink}{rgb}{0.92, 0.3, 0.26}

\usepackage{stackengine}
\stackMath

\definecolor{amethyst}{rgb}{0.6, 0.4, 0.8}
\definecolor{antiquefuchsia}{rgb}{0.57, 0.36, 0.51}
\definecolor{blue-violet}{rgb}{0.54, 0.17, 0.89}

\definecolor{DarkRed}{rgb}{0.6,0,0}
\definecolor{DarkGreen}{rgb}{0,0.6,0}
\definecolor{DarkOrange}{rgb}{1,0.35,0}
\definecolor{DarkBlue}{rgb}{0,0,0.6}

\usepackage{setspace}
\usepackage{authblk}
\usepackage{physics}
\usepackage{multirow}

\usepackage{bbm}

\usepackage{dsfont}

\usepackage{etoolbox}

\let\OLDthebibliography\thebibliography
\renewcommand\thebibliography[1]{
	\OLDthebibliography{#1}
	\setlength{\parskip}{3.0pt plus 2.5pt minus 1.0pt}
	\setlength{\itemsep}{3.0pt plus 2.5pt minus 1.0pt}
}

\usepackage[colorlinks=true,linktocpage=true,linkcolor=DarkOrange,citecolor=DarkGreen,urlcolor=DarkGreen]{hyperref}

\definecolor{goodgreen}{rgb}{0,.6,0.4}

\newcommand{\parity}{\mathcal{H}}

\title{Direct Detection of Dark Baryons \\[0.2em] 
Naturally Suppressed by $\mathcal{H}$-parity \\[1em]}

\author{Pouya Asadi$^1$, Graham D. Kribs$^1$, Chester J. Hamilton Mantel$^2$ \\
{\small \color{purple} 
\texttt{pasadi@uoregon.edu, kribs@uoregon.edu, cmantel@fas.harvard.edu},}
\\\vskip1em
{\small \textit{${}^1$Institute for Fundamental Science and Department of Physics,}\\
\textit{University of Oregon, Eugene, OR 97403, USA} \\\vskip 0.5em
\small \textit{${}^2$Department of Physics, Harvard University, Cambridge, MA 02138, USA}}
}

\date{}

\begin{document}

\maketitle
\begin{abstract}
We identify symmetries in a broad class of vector-like confining dark sectors that forbid the leading electromagnetic moments that would ordinarily mediate dark baryon scattering with the Standard Model. The absence of these operators implies dark baryon dark matter has much smaller cross sections for elastic scattering off nuclei,  leading to suppressed direct detection signals. 
In the confined description, we identify an ``$\parity$-parity'' symmetry that exists in any dark sector with dark quarks transforming under a vector-like representation of a new confining SU($N_c$) gauge theory as well as a vector-like representation of the electroweak group SU(2)$_L$. The parity is independent of $N_c$ and $N_f$, though it is essential that the dark quarks are neutral under hypercharge.
This parity forbids dark hadron electric and magnetic dipole moments, charge radius, and anapole moment, while permitting dimension-7 operators that include polarizability, electroweak loop-induced interactions, as well as lower dimensional electromagnetic \emph{transition} moments between different neutral dark baryon states.
We work out an explicit example, $N_c=N_f=3$, 
that is the most minimal theory with fermionic dark baryons.
In this specific model, we use the non-relativistic quark model to show the magnetic dipole moment and charge radius vanish while the transition moments are non-zero, consistent with $\parity$-parity.
We discuss the implications of a suppressed direct detection signal, emphasizing that this broad class of models provide a well-motivated target for future colliders.
\end{abstract}

\newpage 

\begin{spacing}{1.2}

\tableofcontents

\section{Introduction}

Despite a host of observational evidence for dark matter (DM), its particle nature remains elusive today, see Ref.~\cite{Cirelli:2024ssz} for a recent review. 
An extensive experimental program is underway looking for dark matter's non-gravitational signatures across a range of possible candidates. 
Particle dark matter candidates can arise from a variety of theories that each require understanding the dynamics within the respective dark sectors in order to uncover the signals of the dark sector and the connections with the Standard Model (SM).

Confining dark sectors are among the best-motivated theories with dark matter candidates. 
Just as the confining sector of the SM provides stable matter through the accidental symmetry of baryon number, 
confining dark sectors can also naturally guarantee the stability of dark matter through a conserved dark baryon number. 
Unlike minimal elementary Weakly Interacting Massive Particle (WIMP) models, where the DM mass is typically fixed by obtaining the correct thermal relic abundance, composite theories give rise to much richer dynamics, that can broaden the viable mass range. 
These composite sectors also could provide all the necessary conditions \cite{Sakharov:1967dj} for generating a baryon asymmetry in the universe, and thus explain the similarity of DM and SM abundances today \cite{Bai:2013xga}. 
Composite dark sectors give rise to a plethora of interesting models \cite{Okun:1980mu,Kribs:2009fy,Hambye:2009fg,Bai:2010qg,Fok:2011yc,Frigerio:2012uc,Buckley:2012ky,Bai:2013xga,Antipin:2014qva,Yamanaka:2014pva,Appelquist:2015yfa,Carmona:2015haa,Soni:2016gzf,Harigaya:2016nlg,Dienes:2016vei,DeLuca:2018mzn,Kribs:2018oad,Beylin:2019gtw,Contino:2020god,Morrison:2020yeg,Contino:2020tix}
and novel mechanisms for achieving the dark matter abundance  \cite{Mitridate:2017oky,Acharya:2017szw,Gross:2018zha,Contino:2018crt,Dondi:2019olm,Asadi:2021pwo,Bottaro:2021aal,Baldes:2021aph,Asadi:2022vkc,Carenza:2022pjd,Gouttenoire:2023roe}.
With the search for DM ranging over many fronts, confining dark sectors are one of the most theoretically appealing targets for these experiments, predicting a host of exotic signals at colliders \cite{Strassler:2006im,Han:2007ae,Kang:2008ea,Juknevich:2009ji,Kilic:2009mi,Juknevich:2009gg,Harnik:2011mv,Schwaller:2015gea,Cohen:2015toa,Knapen:2016hky,Knapen:2017kly,Evans:2018jmd,Kribs:2018ilo,Knapen:2021eip,Kuwahara:2023vfc,Batz:2023zef}, 
in direct detection experiments \cite{Bagnasco:1993st,Alves:2009nf,SpierMoreiraAlves:2010err,Bhattacharya:2013kma,Hardy:2015boa}, 
indirect detection \cite{Detmold:2014qqa,Mahbubani:2019pij}, 
and other astrophysical searches \cite{Cline:2013zca,Boddy:2014yra,Krnjaic:2014xza,Buen-Abad:2015ova,Garani:2021zrr}.
They also motivate many lattice studies of dark sectors and lattice simulations of strongly-coupled properties of composite dark matter \cite{LatticeStrongDynamicsLSD:2013elk,LSD:2014obp,Detmold:2014kba,Appelquist:2015yfa,Appelquist:2015zfa,Francis:2018xjd,LatticeStrongDynamics:2020jwi,Brower:2023rqf}.
See Refs.~\cite{Kribs:2016cew,Cacciapaglia:2020kgq,Cline:2021itd,Asadi:2022njl} for recent reviews. 

Constraints from direct detection experiments such as LZ \cite{LZ:2022lsv,LZ2024}, that are searching for nuclear recoil signals of dark matter scattering off nuclei, have put 
remarkably strong limits on the scattering cross section.  
Confining dark sectors, as well as many other thermal dark matter models, can thus be constrained broadly in the mass window of $\mathcal{O}(10)$~GeV to $\mathcal{O}(100)$~TeV.
With no trace of DM interactions to be found in any of these experimental efforts, the lack of evidence motivates exploring theoretical constructions that naturally suppress the DM signal in direct detection searches.  Examples of models that have suppressed elastic scattering cross sections include the pseudo-Dirac Higgsino \cite{Krall:2017xij} in the context of elementary candidates,
and Stealth Dark Matter \cite{Appelquist:2015yfa} in the context of composite candidates from a confining dark sector. 
Such models can not only explain the null-result direct detection searches, but can also guide us toward suitable complementary searches that can fill in the gaps in our experimental endeavors.

In this paper we revisit the direct detection signals of dark matter candidates that arise from confining dark sector theories.  Specifically, we consider a broad class of theories with a set of 
dark fermions that transform in a vector-like representation of a new confining SU($N_c$) gauge group as well as in a vector-like representation of the electroweak SU(2)$_L$ gauge group (with zero hypercharge).  We emphasize that such theories have been studied before in the literature, \textit{e.g.}, see Refs.~\cite{Kilic:2009mi,Antipin:2014qva,Appelquist:2015yfa,Mitridate:2017oky,DeLuca:2018mzn,Kribs:2018ilo,Gross:2018zha}.
The lightest dark baryon, if neutral, will be the dark matter candidate. 
Since the dark quarks transform under SU(2)$_L$, in the confined description the dark hadrons also transform in representations of SU(2)$_L$ (when the confinement scale is above electroweak symmetry breaking) or U(1)$_{\rm em}$ (when the confinement scale is below electroweak symmetry breaking).  For the dark hadron subcomponents that have vanishing SU(2)$_L$ isospin (combined with zero hypercharge), they are both electrically neutral and have no tree-level coupling to the $Z$ boson.  

Below the confinement scale, the generic expectation is that 
the dark hadrons will have a tower of higher dimensional 
operators generated with the SM fields.  
For fermionic dark hadrons, the lowest dimension operators are dimension-5 magnetic and electric dipole moments, as well as the dimension-6 charge radius and anapole moments, see Ref.~\cite{Kavanagh:2018xeh} for a complete list of such operators. 
For bosonic candidates the leading operator is the dimension-6 charge radius.  Contact interactions between dark hadrons
and SM matter are also possible, but since the only interactions
that couple the dark sector to the SM (that we consider) are
SU(2)$_L$ gauge boson exchange, these contact interactions
are captured by ``electroweak loop-induced'' operators
considered in \cite{Essig:2007az,Hisano:2010fy,Hill:2014yka,Hill:2014yxa,Chen:2018uqz,Chen:2023bwg}.

These interactions lead to elastic scattering of dark matter with the SM that can be searched for in direct detection experiments \cite{Pospelov:2000bq,Sigurdson:2004zp,Masso:2009mu,Kribs:2009fy,Barger:2010gv,Banks:2010eh,DelNobile:2012tx,Weiner:2012cb,Weiner:2012gm,DelNobile:2014eta,Appelquist:2015zfa,Kavanagh:2018xeh,Arina:2020mxo,DelNobile:2021wmp,Hambye:2021xvd,PICO:2022ohk,PandaX:2023toi}. 
The low dimensionality of a magnetic (electric) dipole moment
implies that it can have a very large scattering cross section
with nuclei.  If the coefficients of the dipole moments
were $\mathcal{O}(1)$, the current direct detection constraints enforce $m_\chi \gtrsim 100$~TeV \cite{Eby:2023wem} ($m_\chi \gtrsim 10^4$~TeV \cite{Antipin:2014qva}), ruling out a vast swathe of parameter space. 
Such large dark matter masses are not necessarily inconsistent with the abundance mechanisms available to strongly-coupled theories,
but care does need to be taken to avoid annihilation cross sections that violate the unitarity bound \cite{Griest:1989wd,vonHarling:2014kha,Smirnov:2019ngs}.

The goal of this paper is to show that in the vector-like dark sector described above, all of the dimension-5 and dimension-6 electromagnetic moments of neutral dark hadrons vanish. 
The absence of these interactions is due to symmetries:
In the electroweak unbroken phase, the symmetry responsible for this is SU(2)$_L$ itself.  In the electroweak broken phase, we identify a new symmetry called $\parity$-parity that leads to the same 
result.\footnote{The origin of the name, $\parity$-parity, is a matter of some dispute among the authors who have separately advocated for its association with $\parity$odur, $\parity$ades, and $\parity$amilton.  The authors do agree that the resemblance to G-parity (see Section~\ref{subsec:comparison}) made it logical to iterate from $G \rightarrow \parity$.} (In models with some specific values of $N_{c}$, $N_{f}$ closely-related symmetries were imposed or identified, see Refs.~\cite{Kribs:2009fy,Buckley:2012ky}.) 
Our conclusions are broadly applicable to vector-like dark sectors, regardless of the value of $N_{c}$, $N_{f}$, and does not depend on the representation of dark quarks under SU(2)$_L$ or SU($N_c$), nor whether the dark quarks are light or heavy compared to the dark confinement scale. 
We will also work out an explicit example, $N_c=N_f=3$, 
that is the most minimal theory with fermionic dark baryons.
In this specific model, we use the non-relativistic quark model 
\cite{Georgi:1984zwz} to show the magnetic dipole moment and charge radius vanish while the transition moments are non-zero, consistent with $\parity$-parity.

Our results reinforce the importance of collider searches for confining dark sectors that respect $\parity$-parity.  Given that (i) dark hadrons could live within the reach of future colliders,\footnote{See Refs.~\cite{Kang:2008ea,Knapen:2017kly,Kribs:2018ilo,Evans:2018jmd} for some unique signatures of similar models at the LHC.}
(ii) the theoretical appeal of such minimal dark sectors, and (iii) that (we show) the dark baryon dark matter candidate has a highly suppressed  scattering cross section with nuclei, these theories provide a well-motivated target for future high energy collider searches.

The paper is organized as follows. 
We start by introducing the general class of vector-like confinement models that we study in Section~\ref{sec:main}.  We show that 
in the electroweak unbroken phase, SU(2)$_L$ itself is sufficient to suppress the dimension-5 and dimension-6 electromagnetic moments.  
In the electroweak broken phase, 
in Section~\ref{subsec:GPinUV}
we identify a new symmetry called $\parity$-parity, and carefully explain the invariance of the dark sector above and below the confinement scale. 
In Section~\ref{sec:GPandDD} we argue that under mild assumptions about the mass spectrum (no mass degeneracy among the lightest dark baryons), the lack of electromagnetic moments results in a suppression of direct detection signals in such theories, which in turn opens a large parameter space for models that respect $\parity$-parity. 
In Section~\ref{sec:benchmark} we work out an explicit example, $N_c=N_f=3$, 
that is the most minimal theory with fermionic dark baryons.
In this specific model, we use the non-relativistic quark model 
\cite{Georgi:1984zwz} to show the magnetic dipole moment and charge radius vanish while the transition moments are non-zero, consistent with $\parity$-parity.
We conclude in Section~\ref{sec:conclusion}. 
We discuss the exactness of $\parity$-parity given 
interactions with the SM that \emph{do} transform under hypercharge in Appendix~\ref{appx:GPV}, and we consider a slight variation of the parity, $\parity'$-parity, in Appendix~\ref{app:alternativeH}.

\section{Symmetries of Vector-like Confinement}
\label{sec:main}

In this section we introduce the vector-like confinement theories we study in this work and review simple, yet important, facts about the group theory of its gauge symmetries.
The theory space we wish to consider consists of 
a dark sector theory with $N_f$ dark quarks and
anti-quarks transforming as fundamentals of a SU($N_c$) group.\footnote{While we will focus on the fundamental representation of the dark quarks under the new confining gauge group, our results apply to other representations as well.}
For $N_f < \frac{11}{2} N_c$, the SU($N_c$)
group is asymptotically free and confines in the infrared;
we restrict to this theory space throughout the paper.  
The dark quarks transform under the electroweak
group SU(2)$_L$ in the $N_f$-dimensional representation.  For reasons
that will become clear, the quarks are also assumed to be neutral under
U(1)$_Y$.  This is among the most economical models of a new confining
dark sector, where the portal interactions between the dark sector and SM
are through electroweak gauge bosons. Many authors have considered
dark sectors with these ingredients, \textit{e.g.} see Refs.~\cite{Kilic:2009mi,Bai:2010qg,Kribs:2018oad,Kribs:2018ilo,Abe:2024mwa} and references therein.  

In the UV, the renormalizable Lagrangian of the model is 
\begin{eqnarray}
\label{eq:L}
  \mathcal{L} &\supset & -\frac{1}{4} G^{\mu\nu}_aG_{\mu\nu}^a - \theta_\chi \frac{\alpha_\chi}{8\pi} G^{\mu\nu}_a\tilde{G}_{\mu\nu}^a
+ i \mathfrak{q}^\dagger \bar{\sigma}^\mu D_\mu \mathfrak{q} + i \mathfrak{r}^\dagger \bar{\sigma}^\mu D_\mu \mathfrak{r} 
- \overline{m}_0 \left( \mathfrak{q} \mathfrak{r} + h.c. \right),
\end{eqnarray}
where $G^{\mu\nu}_a$ ($\tilde{G}^{\mu\nu}_a$) is the dark gauge group (dual) field strength and $\alpha_\chi \equiv g_\chi^2/(4 \pi)$ is the coupling constant.  The left-handed quarks $\mathfrak{q}$ and $\mathfrak{r}$ transform as shown in Table~\ref{tab:darkquarkstable}.
\begin{table}[]
    \centering
    \begin{tabular}{c|ccc}
                    & SU($N_c$) & SU(2)$_L$  & U(1)$_{\mathfrak{b}}$ \\ \hline
       $\mathfrak{q}$ & $\Box$            & $N_f$ & $1/N_c$ \\ 
       $\mathfrak{r}$ & $\overline{\Box}$ & $N_f$ & $-1/N_c$ \\ 
    \end{tabular}
    \caption{Left-handed dark quark quantum numbers under the confining dark gauge group SU($N_c$), the SM electroweak gauge group SU(2)$_L$, and the dark baryon number U(1)$_\mathfrak{b}$. Aside from the dark gauge fields, these are the only new fields in our theory.}
    \label{tab:darkquarkstable}
\end{table}
The global flavor symmetry of quarks, $U(N_f)_L \times U(N_f)_R$, is
both spontaneously broken by SU($N_c$) confinement, as well as 
explicitly broken by the quark mass $\overline{m}_0$ down to
SU($N_f$)~$\times$~U(1)$_{\mathfrak{b}}$, where U(1)$_{\mathfrak{b}}$ denotes the dark baryon number symmetry. 
We will also utilize the 4-component Dirac fermion notation
\begin{eqnarray}
\mathbf{Q} \equiv { \mathfrak{q} \choose \mathfrak{r}^\dagger },
\end{eqnarray}
that allows the kinetic and mass terms to be written simply as
\begin{eqnarray}
  i \bar{\mathbf{Q}} \slashed{D} \mathbf{Q}
  - \overline{m}_0 \bar{\mathbf{Q}} \mathbf{Q} \, .
\end{eqnarray}
The diagonal subgroup flavor symmetry SU($N_f$) is further explicitly
broken by the gauging of the electroweak interactions
for $N_f > 2$. 
We will mainly consider a strong sector that preserves CP,
with $\theta_\chi=0$, postponing implications of non-zero $\theta_\chi$ to future work.
The lightest baryon is stabilized by the accidental symmetry
U(1)$_{\mathfrak{b}}$,
\textit{i.e.} dark baryon number, in close analogy to baryon number of the SM\@.  After electroweak symmetry breaking, so long as the lightest dark baryon is neutral under the electromagnetism, it can constitute a viable DM candidate.
In what follows we review further basic facts about the model in two qualitatively different limits.

\subsection{Vector-like confinement with $SU(2)_L$:
$\Lambda_d > v_h$ limit}
\label{sec:Lambdaabovev}

Consider first the case where the SU($N_c$)
dark sector theory confines \emph{before} 
electroweak symmetry breaking, $\Lambda_d > v_h$ with $\Lambda_d$ denoting the dark confinement scale and $v_h \simeq 246$~GeV 
the Higgs vacuum expectation value.
In the effective field theory (EFT) below $\Lambda_d$,
the theory consists of dark color-neutral 
baryons and mesons that transform as complete
representations under SU(2)$_L$.  An extensive 
analysis of the set of representations that arise
for a given UV theory, and the spectrum that
is expected, will be presented in
Ref.~\cite{ABK}.  With a small number of colors
and flavors, it is straightforward to identify
the SU(2)$_L$ representations, and we present
an explicit example in Section~\ref{sec:benchmark}.

Thus, we have a set of hadrons that transform in 
set of $\left\{ N_i, N_j, \ldots \right\}$ 
representations under SU(2)$_L$.
It is straightforward to write the 
electroweak interactions
of the hadrons that arise from their kinetic terms.
Alas, the relevant effective theory for baryons (and heavy mesons) 
is valid only in the non-relativistic limit. 
In order to simplify the presentation, 
for the spin-0 and spin-1/2 hadrons, 
we will use relativistic field notation and normalization, recognizing that the 
non-relativistic limit is to be taken when 
actually using the effective theory of these baryons.

Hadrons in the $N$ representation 
of SU(2)$_L$ have $N$
states that can be classified by their $T_3^a$ 
quantum numbers, 
\begin{equation}
\begin{array}{rcl}
    \mbox{even} \; N: & & T_3^a(B_a) \;  
    = \; \big\{ \; - (N-1)/2, \, \ldots, \, -1/2, \, 1/2, \, \ldots, \, (N-1)/2 \; \big\}, \\ 
    \mbox{odd} \; N: & & T_3^a(B_a) \;  
    = \; \big\{ \; - (N - 1)/2, \, \ldots, \, -1 \, , 0, \, 1, \, \ldots, \, (N - 1)/2 \; \big\} \, .
\end{array}
\label{eq:T3quantumnumbers}
\end{equation}
The index $a$ will therefore be taken to iterate
over the $T_3^a$ quantum numbers, 
i.e., $a = -(N-1)/2, \ldots, (N-1)/2$.
Since the dark quarks do not transform under
hypercharge, neither do the hadrons.  
Hence, after electroweak symmetry breaking,
hadrons within an
$N$-dimensional SU(2)$_L$ representation become 
simply $N$ distinct hadrons with electric charges
given by $Q_a = T_3^a$.
For every hadron in an odd 
SU(2)$_L$ representation, 
there is one electrically neutral hadron
($a=0$).

Consider now a baryon state - and for our
purposes we will only consider the smallest
spin representation available under
SU($N_c$).  For even $N_c$, this is a 
complex scalar baryon $\phi_{B}$;
for odd $N_c$ this is a spin-1/2 baryon
that we denote by $\chi_B$ 
(a four component Dirac spinor).
For odd $N_c$, at the renormalizable level, 
the baryon kinetic term includes the interaction
with electroweak gauge bosons
\begin{eqnarray}
   g \, \bar{\chi}_{B,a} \, \gamma^\mu J_{ab}^i \, \chi_{B,b} \, W_\mu^i
\label{eq:Brenormalizable}
\end{eqnarray}
where $i=1,2,3$ and $J_{ab}^i$ is the SU(2)$_L$
group matrix for the $N$ representation.
For odd $N$, the specific state $\chi_{B,0}$ has
zero electric charge.  It has no 
renormalizable couplings to the photon or
the $Z$ since it does not transform under
hypercharge and the interaction with $W^3$
vanishes due to SU(2)$_L$ group theory.
Specifically the group matrix $J_{ab}^3$
is diagonal with the form
\begin{eqnarray}
    J_{ab}^3 &=& \left( 
    \begin{array}{ccccc} 
    (N -1)/2 &        &   & & \\
             & \ddots &   & & \\
             &        & 0 & & \\
             &        &   & \ddots & \\
             &        &   &        & -(N-1)/2 
    \end{array} \right) \quad \qquad (\mbox{for odd} \; N) \, .
\end{eqnarray}

Higher dimensional interactions of fermionic baryons include 
\begin{eqnarray}
\frac{1}{\Lambda_d} \;
\bar{\chi}_{B,a} \, \sigma^{\mu\nu} (\gamma^5) \, J_{ab}^i \, \chi_{B,b} \, F_{W,\mu\nu}^i, \label{eq:dim5fluv} \\
\frac{1}{\Lambda_d^2} \;
\bar{\chi}_{B,a} \, \gamma^\mu (\gamma^5)\, J_{ab}^i \, \chi_{B,b} \, \partial^\nu F_{W,\mu\nu}^i,
\end{eqnarray}
and for spin-0 baryons,  
at dimension-6 we can write 
\begin{equation}
\frac{1}{\Lambda_d^2} \;
\phi_{B,a}^\dagger \, \overset{\text{\tiny{$\longleftrightarrow$}}}{\partial^{\mu}} \, J_{ab}^i \, \phi_{B,b} \, \partial^\nu  F_{W,\mu\nu}^i \, .
\label{eq:dim6sluv}
\end{equation}
These operators involving the SU(2)$_L$ 
field strength have the \emph{same} SU(2)$_L$
group elements, namely $J_{ab}^i$, as the renomralizable term
Eq.~(\ref{eq:Brenormalizable}). 
Consequently, for odd $N_c$, 
the neutral spin-1/2 baryon
$\chi_{B,0}$
and neutral scalar baryon $\phi_{B,0}$
have no dimension-5 or dimension-6 interactions
with $F^3_{W,\mu\nu}$.  Since the dark
sector does not transform under hypercharge,
higher dimensional interactions with 
$F_{B,\mu\nu}$ are not generated after
confinement (considering only interactions 
from within the dark sector). 
Hence, in the effective theory below
electroweak symmetry breaking,
no dimension-5 or dimension-6
operators involving a single field strength
of the photon 
$F_{\mu\nu} = c_w F_{W,\mu\nu}^3 + s_w F_{B,\mu\nu}$
are generated.
This means that dimension-5 and dimension-6 electromagnetic moments are \emph{not generated}
for all electrically neutral hadrons that 
arise from the dark sector of 
Eq.~(\ref{eq:L}).

\subsection{Vector-like confinement with $\Lambda_d < v_h$}
\label{sec:lambdabelowv}

If the dark sector confinement scale is below 
electroweak symmetry breaking, $\Lambda_d < v_h$,
the $W$ and $Z$ bosons acquire mass,
are integrated out, resulting in an effective theory
with dark quarks that transform only under
U(1)$_{\rm em}$.  In this regime, the effective theory 
consists of $N_f$ dark quarks $\mathfrak{q}_a$
($\mathfrak{r}_a$)
in the (anti-)fundamental 
representation of SU($N_c$).
For odd $N_f$, the neutral quarks
$\mathfrak{q}_0$, $\mathfrak{r}_0$ have no renormalizable interactions
with electromagnetism.

Once we transition into the confined theory,
at scales below $\Lambda_d$ $(< v_h)$, 
we have a set of dark color-neutral hadrons 
with electric charges.  Performing the same
analysis as in Section~(\ref{sec:Lambdaabovev}),
we expect higher dimension operators
for dark fermionic baryons to be of the form
\begin{eqnarray}
\frac{c_{5,a}}{\Lambda_d} \;
\bar{\chi}_{B,a} \, \sigma^{\mu\nu} (\gamma^5) \, \chi_{B,a} \, F_{\mu\nu}, \label{eq:dim5fllv} \\
\frac{c_{6,a}}{\Lambda_d^2} \;
\bar{\chi}_{B,a} \, \gamma^\mu (\gamma^5)\, Q_a \, \chi_{B,a} \, \partial^\nu F_{\mu\nu},
\end{eqnarray}
and for dark scalar baryons to be
\begin{equation}
\frac{c'_{6,a}}{\Lambda_d^2} \;
\phi_{B,a}^\dagger \, \overset{\text{\tiny{$\longleftrightarrow$}}}{\partial^{\mu}} \, \phi_{B,a} \, \partial^\nu  F_{\mu\nu} \, ,
\label{eq:dim6sllv}
\end{equation}
in terms of the field strength of the photon $F_{\mu\nu}$ that is the only remaining 
massless gauge boson in the EFT\@. 
The absence of the full SU(2)$_L$
gauge symmetry means the higher dimensional
operators in
Eqs.~(\ref{eq:dim5fllv})-(\ref{eq:dim6sllv}) 
do not have SU(2)$_L$ symmetry constraints
on the structure of the interactions
that were present for 
Eqs.~(\ref{eq:dim5fluv})-(\ref{eq:dim6sluv}).
In particular, we have lost the 
SU(2)$_L$ structure $J_{ab}^i$ in favor of  
otherwise arbitrary coefficients 
$c_{5,a}$, $c_{6,a}$, and $c'_{6,a}$.

\section{$\parity$-Parity in Vector-like Confinement}
\label{subsec:GPinUV}

In this section we introduce the central concept of this paper -- $\parity$-parity -- and show that the Lagrangian in Eq.~\eqref{eq:L} automatically contains this discrete global symmetry in addition to the gauge symmetries that are already manifest.
As we will see, $\parity$-parity
is preserved after electroweak symmetry 
breaking, providing further constraints on the
operators in
Eqs.~(\ref{eq:dim5fllv})-(\ref{eq:dim6sllv}).

In the dark quark theory above electroweak symmetry breaking, 
$\mathfrak{q}$ and $\mathfrak{r}$ transform under the same (pseudo)real representation $N_f$ of SU($2$)$_L$.  
In addition to the local SU(2)$_L$ gauge symmetry, Eq.~(\ref{eq:L}) also admits a global parity transformation in combination with charge conjugation of the electroweak gauge bosons.
Consider the following transformation
acting on components of each SU(2)$_L$ multiplet of dark quarks 
\begin{eqnarray}
\label{eq:PfUV}
\mathcal{S} : \left\{
\begin{array}{lcr}
\mathfrak{q}_{a} & \rightarrow & S_{ab} \mathfrak{q}_{b} \\ 
\mathfrak{r}_{a} & \rightarrow & \mathfrak{r}_{b} S_{ba}^\dagger 
\end{array} \right. 
\end{eqnarray}
in two-component notation, or equivalently
\begin{eqnarray}
\label{eq:PfUVQ}
\mathcal{S} : 
\begin{array}{lcr}
\textbf{Q}_{a} & \rightarrow & S_{ab} \textbf{Q}_{b} 
\end{array} 
\end{eqnarray}
in four-component notation.
The $\mathcal{S}$ transformation acts as a parity, 
\begin{eqnarray}
S &=& \mathrm{exp} \left( i \pi J_2 \right) = (-1)^{T_3^a+\mathtt{k}} \delta_{T_3^a,-T_3^b} \, 
\label{eq:expipij2}
\end{eqnarray}
in terms of the $T_3^a$ quantum numbers for the $a^{\rm th}$ component of the $N_f$ representation.
We use the same notation as Eq.~(\ref{eq:T3quantumnumbers})
with $a,b$ running over the $T_3$ values, with $\mathtt{k}=N_f/2 + 1/2$ ($\mathtt{k}=N_f/2 - 1/2$) for even (odd) $N_f$.\footnote{For (even) odd values of $N_f$, the quarks transform in a (pseudo)real representation of SU(2)$_L$; as a result of this, we find $S^\dagger = \pm S$, with the sign appearing for pseudoreal representations (even $N_f$ values).} 
The $\mathcal{S}$ operation is a \emph{global} 
$\pi$ rotation of the fields implemented with the SU($2$)$_L$ generator $J_2$.  
After electroweak symmetry breaking, the electric charges of the component fields are identified as simply $Q_a = T_3^a$.  
For example, for $N_f=(3,4,5)$, $\mathcal{S}$ acts on $\mathfrak{q}$ dark quarks as 
\begin{equation}
N_f=3:  ~~~\mathcal{S} :
\left( \begin{matrix}
\mathfrak{q}_{1} \\ \mathfrak{q}_{0} \\ \mathfrak{q}_{-1}    
\end{matrix}
\right) \rightarrow 
\left( \begin{matrix}
0 & 0 & 1 \\
0 & -1 & 0 \\
1 & 0 & 0 
\end{matrix}
\right)
\left( \begin{matrix}
\mathfrak{q}_{1} \\ \mathfrak{q}_{0} \\ \mathfrak{q}_{-1}    
\end{matrix}
\right)
= 
\left( \begin{matrix}
\mathfrak{q}_{-1} \\ -\mathfrak{q}_{0} \\ \mathfrak{q}_{1}    
\end{matrix}
\right), \label{eq:exampleG3}
\end{equation}
\begin{equation}
N_f=4: ~~~\mathcal{S} :
\left( \begin{matrix}
\mathfrak{q}_{3/2} \\ \mathfrak{q}_{1/2} \\ \mathfrak{q}_{-1/2} \\ \mathfrak{q}_{-3/2}    
\end{matrix}
\right) \rightarrow 
\left( \begin{matrix}
0 & 0 & 0 & 1 \\
0 & 0 & -1 & 0 \\
0 & 1 & 0 & 0 \\
-1 & 0 & 0 & 0 
\end{matrix}
\right)
\left( \begin{matrix}
\mathfrak{q}_{3/2} \\ \mathfrak{q}_{1/2} \\ \mathfrak{q}_{-1/2} \\ \mathfrak{q}_{-3/2}   
\end{matrix}
\right)
= 
\left( \begin{matrix}
\mathfrak{q}_{-3/2} \\ -\mathfrak{q}_{-1/2} \\ \mathfrak{q}_{1/2} \\ -\mathfrak{q}_{3/2} 
\end{matrix}
\right),
    \label{eq:exampleG4}
\end{equation}
\begin{equation}
N_f=5:  ~~~\mathcal{S} :
\left( \begin{matrix}
\mathfrak{q}_{2} \\ \mathfrak{q}_{1} \\ \mathfrak{q}_{0} \\ \mathfrak{q}_{-1} \\ \mathfrak{q}_{-2}    
\end{matrix}
\right) \rightarrow 
\left( \begin{matrix}
0 & 0 & 0 & 0 & 1 \\
0 & 0 & 0 & -1 & 0 \\
0 & 0 & 1 & 0 & 0 \\
0 & -1 & 0 & 0 & 0 \\
1 & 0 & 0 & 0 & 0 
\end{matrix}
\right)
\left( \begin{matrix}
\mathfrak{q}_{2} \\ \mathfrak{q}_{1} \\ \mathfrak{q}_{0} \\ \mathfrak{q}_{-1} \\ \mathfrak{q}_{-2}   
\end{matrix}
\right)
= 
\left( \begin{matrix}
\mathfrak{q}_{-2} \\ -\mathfrak{q}_{-1} \\ \mathfrak{q}_{0} \\ -\mathfrak{q}_{1} \\ \mathfrak{q}_{2} 
\end{matrix}
\right),
    \label{eq:exampleG5}
\end{equation}
where subscripts denote the $T_3^a$ (equal to $Q_a$) quantum numbers of the component fields.  Identical equations are applicable to $\mathfrak{r}$ quarks as well.
It is worth noting that the neutral
dark quark $\mathfrak{q}_0$ has different eigenvalues under $\mathcal{S}$ for different $N_f$-multiplets,
namely $-1$ ($+1$) when $(N_f - 1)/2$ is odd (even).

Performing a global parity $\mathcal{S}$ operation on all of the dark quarks leaves the Lagrangian trivially invariant \emph{except} for the interaction with SU(2)$_L$ gauge bosons in the covariant derivative. For any representation of the dark quarks ($\mathfrak{q}$,~$\mathfrak{r}$)
under SU(2)$_L$, the $\mathcal{S}$ transformation results in
\begin{eqnarray}
    \label{eq:SonJ}
    S_{ca}^\dagger J_{ab}^i S_{bd} & = & 
    - \left[J_{cd}^i\right]^T \; = \; 
    \left\lbrace \begin{array}{ccc}
        J_{cd}^i & \; & i=2 \\
        -J_{cd}^i & \; & i=1,3 \, 
    \end{array} \right.  ,
\end{eqnarray}
where $\left[J_{cd}^i\right]^T = J_{dc}^i$, 
and this can be equivalently written as 
\begin{eqnarray}
    \label{eq:SonJ2}
    S^\dagger \left( \begin{array}{c}
                      J^+ \\
                      J^3 \\
                      J^- \end{array} \right) S & = & 
    - \left( \begin{array}{c}
                      J^- \\
                      J^3 \\
                      J^+ \end{array} \right) \, ,
\end{eqnarray}
where $J^\pm = (J^1 \pm i J^2)/\sqrt{2}$. 
This is obviously not a symmetry of the dark quark SU(2)$_L$ interaction terms by itself.  Nevertheless, 
the sign flip and transpose resulting from
Eq.~(\ref{eq:SonJ}) can be compensated
by simultaneously performing charge conjugation to the SU(2)$_L$ gauge bosons\footnote{In any SU($N$) Yang-Mills theory, guage fields associated with (real-) complex-valued generators remain unchanged (get a sign) under charge conjugation. We thank Spencer Chang for elaborating on the action of charge conjugation on gauge fields.}
\begin{eqnarray}
W^c_{\mu,i} \equiv \mathcal{C} W_{\mu,i} \mathcal{C} & = & \left\lbrace \begin{matrix}
        W_{\mu,i} & i=2 \\
        -W_{\mu,i} & i=1,3 \, \,
    \end{matrix}\right. \label{eq:Wchargeconjugation}
\end{eqnarray}
or equivalently
\begin{eqnarray}
    \label{eq:Wchargeconjugation2}
    \mathcal{C} \left( \begin{array}{c}
                      W_\mu^+ \\
                      W_\mu^3 \\
                      W_\mu^- \end{array} \right) \mathcal{C} & = & 
    - \left( \begin{array}{c}
                      W_\mu^- \\
                      W_\mu^3 \\
                      W_\mu^+ \end{array} \right) \, .
\end{eqnarray}
Thus, we see that this dark sector theory 
contains an automatic parity symmetry -- $\parity$-parity -- that we denote by 
\begin{eqnarray}
\parity \equiv \mathcal{S} \otimes \mathcal{C}
\end{eqnarray}
where $\mathcal{S}$ acts on all fields transforming under SU(2)$_L$, and charge conjugation $\mathcal{C}$ acts on the electroweak gauge bosons.
For a 2-component Weyl field 
$\mathfrak{f}_a$ in an 
$N$ dimensional SU(2)$_L$ representation,
\begin{eqnarray}
    \parity \, \left[ \, i \, \mathfrak{f}_a^\dagger \bar{\sigma}^\mu J_{ab}^i \mathfrak{f}_b W_\mu^i \right] & \rightarrow & i \, \mathfrak{f}_c^\dagger S_{ca}^\dagger \bar{\sigma}^\mu J_{ab}^i S_{bd} \mathfrak{f}_d \, \mathcal{C} W_\mu^i \mathcal{C} 
    \label{eq:Hsimtwocomponent} \\
    & = & i \, \mathfrak{f}_c^\dagger \bar{\sigma}^\mu \left( S_{ca}^\dagger J_{ab}^i S_{bd} \right) \mathfrak{f}_d (W_\mu^i)^c \nonumber \\
    & = & i \, \mathfrak{f}_c^\dagger \bar{\sigma}^\mu J_{cd}^i \mathfrak{f}_d W_\mu^i \, . \nonumber 
\end{eqnarray}
Hence, the kinetic terms of the dark sector
Lagrangian, Eq.~(\ref{eq:L}), are invariant.
A key observation in this paper is that 
dark sector interactions preserve $\parity$-parity 
for any value of $N_{c,f} \ge 2$.

$\parity$-parity also acts on the SU(2)$_L$
multiplets of the SM\@.  
The SU(2)$_L$ gauge boson
interaction terms of 
SM quark doublets, lepton doublets,
and Higgs doublets are obviously 
also invariant, see Appendix~\ref{appx:GPV}.
Here it is useful to emphasize $\parity$-parity
transforms left-handed fermions 
in SU(2)$_L$ representations into 
left-handed fermions,
as is evident in Eq.~(\ref{eq:Hsimtwocomponent}). 
We take $\parity$-parity act trivially on 
dark color interactions for the dark sector
fermions as well as SU(3)$_c$ interactions for the SM quarks.  

The remaining outstanding issue is how 
$\parity$-parity acts on the hypercharge
gauge boson $B_\mu$.  However, the choice
cannot be separated from what happens to
$W_\mu^i$, since electroweak symmetry breaking
causes SU(2)$_L \; \times \; $U(1)$_B \; \rightarrow \; $U(1)$_{\rm em}$.
Since the photon is a linear combination of
$W_\mu^3$ and the hypercharge gauge boson $B_\mu$,
\textit{i.e.} $A_\mu = s_w W_\mu^3 + c_w B_\mu$, 
there is only one way for how $\parity$-parity could act on $B_\mu$ such that its action on $A_\mu$ is the same as charge conjugation.
Namely, we require $\parity$-parity to act on $B_\mu$ as a charge conjugation operator 
\begin{eqnarray}
    B_\mu^c \; \equiv \; \mathcal{C} B_\mu \mathcal{C} 
&=& -B_\mu \, . \label{eq:Bchargeconjugation}
\end{eqnarray}
Then, below the electroweak breaking scale, 
since the SU(3)$_c \; \times \; $U(1)$_{\rm em}$ interactions are invariant under charge conjugation, in this low energy effective theory there are no interactions that violate the $\parity$-parity.  

What is clear, however, is that the SM quark
doublet, lepton doublet, and Higgs doublet
interactions with hypercharge violate 
$\parity$-parity.  This is because the
$\mathcal{S}$ operation on the SM fermions current
is trivial, and yet the $\mathcal{C}$
operation causes 
$B_\mu \rightarrow B_\mu^c = - B_\mu$.
So under $\parity$-parity, the hypercharge 
terms change sign.  Nevertheless,
as we discuss in Appendix~\ref{appx:GPV},
hypercharge-induced $\parity$-parity 
violation leads to phenomenologically-negligible changes to
our results, and so we ignore $\parity$-parity violation for the remainder of this
discussion. Then, in Appendix~\ref{app:alternativeH}, we also comment on an alternative interpretation
of the same operators involving $F_{B,\mu\nu}$, where the would-be $\parity$-parity violating contribution is reinterpreted in terms of a different parity, $\parity'$-parity, that does not transform $B_\mu$.

\subsection{$\parity$-Parity in the Confined Phase}
\label{subsec:GPinIR}

Let us now study implications of (exact) $\parity$-parity in the confined phase of the dark sector. 
It is clear from Eq.~(\ref{eq:Hsimtwocomponent})
that both the $\mathfrak{q}$ and $\mathfrak{r}$
interaction terms are invariant.
Confinement of the dark sector leads to the dark quark condensate 
\begin{eqnarray}
    \langle \mathrm{tr}(\mathfrak{r} \mathfrak{q} + h.c.) \rangle & \sim & \Lambda_d^3 \, ,
\end{eqnarray}
which under $\parity$-parity,
\begin{eqnarray}
    \mathcal{H} \, \left[ \langle \mathrm{tr}(\mathfrak{r} \mathfrak{q} + h.c.) \rangle \right] & \rightarrow & 
    \langle \mathrm{tr}(\mathfrak{r} S^\dagger S \mathfrak{q}  + h.c. ) \rangle \; = \; 
    \langle \mathrm{tr}(\mathfrak{r} \mathfrak{q} + h.c.) \rangle \, ,
\end{eqnarray}
is invariant.  This follows because the field content of the dark sector is vector-like,
which is the crucial ingredient. 
All of this is under the assumption of $\theta_\chi = 0$, and thus $\parity$-parity acts as a vector-like symmetry that is well-known to remain unbroken after confinement \cite{Vafa:1983tf}.
In the confined phase, dark quarks are bound together to form the new degrees of freedom of the theory, namely dark hadrons.  
Since $\parity$-parity is preserved after confinement, the interactions of dark hadrons also respect $\parity$-parity.

In the confined description where $\Lambda_d < v_h$,
the dark hadrons are classified by their electric charges.  $\parity$-parity acts on the dark hadrons exactly as given in Eq.~(\ref{eq:expipij2}), with $T_3^a$ replaced by the electric charge $Q_a$ of the dark hadron.  With this replacement, the $S$ transformation on dark hadrons is
\begin{equation}
    S_{ab} | B_b \rangle = \pm | B_a \rangle \, .
    \label{eq:PfonBischarged}
\end{equation}
Electrically charged dark hadrons are not eigenstates of $\parity$-parity, since $\mathcal{S}$ transforms a dark baryon $|B_b\rangle$ with charge $Q_b$ 
to $|B_a\rangle$ with charge $Q_a = -Q_b$.  
This is unsurprising since electromagnetic interactions are $\parity$-parity invariant:  namely, the $\mathcal{S}$ transformation flips a dark hadron with charge $Q_b$ to the opposite charge $Q_a=-Q_b$, and simultaneously charge conjugates the photon field, $A_\mu \rightarrow A_\mu^c = -A_\mu$, leaving all of the renormalizable dark hadron interaction terms with the photon invariant.  

The one critically important special case is when there is an electrically neutral state, \textit{i.e.}, $a=b=0$, 
\begin{equation}
    S_{00} | B_0 \rangle = \pm | B_0 \rangle \, .
    \label{eq:PfonBis}
\end{equation}
where the eigenvalue is $\pm 1$ depending on in which representation under SU(2)$_L$ it was embedded.  This global parity remnant of SU(2)$_L$, where neutral dark hadrons are eigenstates of $\parity$-parity with $\pm 1$ eigenvalues plays an important role in understanding which higher dimensional operators are permitted, and which ones are forbidden.

\subsection{Transformation of Higher Dimensional  Operators under $\parity$-Parity}
\label{subsec:operatorstrans}

In the low energy EFT below electroweak symmetry breaking, we can now categorize all effective interactions of dark baryons, focusing on the neutral baryons, $B_0$. 
(Phenomenology of effective operators have been studied in the literature extensively \cite{Pospelov:2000bq,Sigurdson:2004zp,Masso:2009mu,Barger:2010gv,Banks:2010eh,DelNobile:2012tx,Weiner:2012cb,Weiner:2012gm,DelNobile:2014eta,Ovanesyan:2014fha,Appelquist:2015zfa,Kavanagh:2018xeh,Arina:2020mxo,Hambye:2021xvd,PICO:2022ohk,PandaX:2023toi}.)
As we have emphasized, $\parity$ acts as charge conjugation on $A^\mu$, and thus the photon field strength transforms as
\begin{equation}
\parity:~ F^{\mu\nu} \rightarrow \left( F^{\mu\nu} \right)^c \; = \; - F^{\mu\nu} \, .
    \label{eq:exampleGFmunu}
\end{equation}
Importantly, Eqs.~\eqref{eq:PfonBis} and \eqref{eq:exampleGFmunu} imply all dimension-5 and dimension-6 operators that involve the same neutral dark baryon coupled to one power of $F_{\mu\nu}$ are \emph{odd} under $\parity$.
Again, we present the operators in terms of a fermionic baryon $\chi_{B,0}$ or scalar baryon $\phi_{B,0}$ fields, and for ease of presentation, we use the
relativistic normalization of the fields.

The $\parity$-odd operators for fermionic baryons are the magnetic and the electric dipole moments,
\begin{equation}
 \parity: \bar{\chi}_{B,0} \sigma^{\mu\nu} (\gamma^5) \chi_{B,0} F_{\mu\nu} \;\rightarrow\; - \, \bar{\chi}_{B,0} \sigma^{\mu\nu} (\gamma^5) \chi_{B,0} F_{\mu\nu} \, ,
    \label{eq:PfIRMDM}
\end{equation}
as well as the dimension-6 charge radius and anapole moment
\begin{equation}
 \parity: \bar{\chi}_{B,0} \gamma^\mu (\gamma^5)\chi_{B,0} \partial^\nu F_{\mu\nu} \;\rightarrow\; - \, \bar{\chi}_{B,0} \gamma^\mu (\gamma^5) \chi_{B,0} \partial^\nu F_{\mu\nu} \, ,
    \label{eq:PfIRCR}
\end{equation}
and for spin-0 baryons the dimension-6 charge radius\footnote{In the non-relativistic limit $\phi_{B,0}$ is a dimension-3/2 field and the derivative is replaced with the DM velocity \cite{Bagnasco:1993st}, leaving the operator dimension unchanged.}
\begin{equation}
 \parity:  \phi_{B,0}^\dagger \overset{\text{\tiny{$\longleftrightarrow$}}}{\partial^{\mu}} \phi_{B,0} \partial^\nu  F_{\mu\nu}   \;\rightarrow\; - \, \phi_{B,0}^\dagger \overset{\text{\tiny{$\longleftrightarrow$}}}{\partial^{\mu}} \phi_{B,0} \partial^\nu  F_{\mu\nu} \, .
    \label{eq:PfIRCRboson}
\end{equation}
Since the Lagrangian preserves $\parity$ in the UV and in the IR (when $\theta_\chi=0$), neutral mass eigenstate baryons 
do not have magnetic or electric dipole moments, a charge radius, or an anapole moment.  (Of course, when $\theta_\chi=0$, CP conservation also forbids electric dipole and anapole moments.)
This is the main implication of $\parity$-parity.

At dimension-7, there are a plethora of higher dimensional operators involving neutral baryons that are permitted by $\parity$-parity. These include several operators that are simple bilinears of the baryons, such as dark baryon polarizability (see Ref.~\cite{Kavanagh:2018xeh} for a full list of polarizability operators)
\begin{equation}
 \parity:  \bar{\chi}_{B,0} \chi_{B,0} F_{\mu\nu} F^{\mu\nu}    \;\rightarrow\; + \, \bar{\chi}_{B,0} \chi_{B,0} F_{\mu\nu} F^{\mu\nu}  \, .
    \label{eq:polarizability}
\end{equation}
There is also a polarizability for scalar baryons that becomes dimension-7 in the non-relativistic limit. 
Finally, ``electroweak loop-induced'' operators arise after integrating out massive electroweak gauge bosons and the Higgs boson \cite{Essig:2007az,Hisano:2010fy,Hill:2014yka,Hill:2014yxa,Chen:2018uqz,Chen:2023bwg}.  
In the non-relativistic limit, these are matched onto
\begin{equation}
  \bar{\chi}_{B,0} \chi_{B,0} \left( \sum_q \left[ c_q^{(0)} O_q^{(0)} + c_q^{(2)} v_\mu v_\nu O_q^{(2)\mu\nu} \right] \right) \, ,
    \label{eq:EWloop}
\end{equation}
where the twist-0 and twist-2 contributions are, respectively, 
\begin{equation}
    O_q^{(0)} = m_q \bar{q}q,~~~ O_q^{(2)\mu\nu} = \frac{1}{2} \bar{q} \left(  \gamma^{\lbrace \mu} i D_-^{\nu \rbrace} - \frac{1}{d} g^{\mu\nu} i \slashed{D}_- \right) q,
    \label{eq:defOqs}
\end{equation}
with $D_- = \overrightarrow{D} - \overleftarrow{D}$, $A^{\lbrace \mu } B^{\nu \rbrace} = (A^\mu B^\nu + A^\nu B^\mu)/2$, and $d=4-2\epsilon$ is the spacetime dimension when dimensional regularization is employed, see Ref.~\cite{Hill:2014yka} for further details and Table~\ref{tab:summary_bounds} for the relevant diagrams. 
There are also dimension-7 electroweak loop-induced operators that match onto the gluon field strength squared (called $O_g^{(0)}$ and $O_g^{(2)}$ in Ref.~\cite{Hill:2014yka}) that receive two-loop contributions and are subleading to $O_q^{(0)}$ and $O_q^{(2)\mu\nu}$. 
The electroweak contributions in Eq.~\eqref{eq:EWloop} will turn out to be the leading contributions to the cross section for direct detection as we explain in detail in Section~\ref{sec:GPandDD}.\footnote{Assuming $B_0$ is a not a pure singlet under SU(2)$_L$; otherwise the coefficients of higher dimensional operators that arise from electroweak loop-induced contributions will vanish. We will explore this intriguing possibility in Ref.~\cite{ABK}.}

Finally, there is one last operator that is permitted in the effective field theory, the dimension-5 interaction with the Higgs doublet
\begin{eqnarray}
 \parity:  \bar{\chi}_{B,0} \chi_{B,0} H^\dagger H
 &\rightarrow& + \, \bar{\chi}_{B,0} \chi_{B,0} H^\dagger H .
    \label{eq:baryondim5higgs}
\end{eqnarray}
This operator coefficient vanishes in the case where we include only renormalizable interactions in the Lagrangian, \textit{i.e.} Eq.~(\ref{eq:L}).  If, however, we extended Eq.~(\ref{eq:L}) to include the dimension-5 interaction $\bar{\mathbf{Q}} \mathbf{Q} H^\dagger H$, this would lead to a non-zero coefficient proportional to the strength of the operator in Eq.~(\ref{eq:baryondim5higgs}), which can be captured by
the low energy effective theory operators of Eq.~(\ref{eq:EWloop}) \cite{Chen:2018uqz}. Given that the existence of this operator requires yet higher energy physics to UV complete the interaction, we will take the dark sector Lagrangian to be precisely Eq.~(\ref{eq:L}), and thus the dimension-5 Higgs operator Eq.~(\ref{eq:baryondim5higgs}) is not generated.


\subsection{Comments on Other Parity Symmetries of Confining Dark Sectors}
\label{subsec:comparison}

In this section we comment on a few closely related parities of confining dark sectors and their relationship to $\parity$-parity. 
First, it is interesting to consider what would happen if
the dark quarks transformed under just U(1)$_Y$, instead of SU(2)$_L$,
with the hypercharge of each dark quark equal to the $T_3$ eigenvalue
of SU(2)$_L$ (so that their electric charges under
U(1)$_{\rm em}$ match precisely).  All of the results concerning
the interactions of the dark quarks with U(1)$_{\rm em}$ remain.
However, the dark quark mass terms need no longer be SU(2)$_L$ symmetric, in general, and thus $\parity$ can be preserved only if
the dark quark masses \emph{accidentially} preserve an
SU(2) global flavor symmetry.  This is what was proposed in
Quirky Dark Matter \cite{Kribs:2009fy}, under the name
``ud-parity.''  In that work, ud-parity was \emph{imposed} on
the (right-handed) fermion sector of that theory, leading to an
accidental $\parity$-parity on the quirky dark baryons. (See also Ref.~\cite{Buckley:2012ky} for a discussion of the same ud-parity in the small dark quark mass limit.)

An identical Lagrangian to our Eq.~\eqref{eq:L} was considered in Ref.~\cite{Bai:2010qg} as well. 
They noticed that
the action of $\mathcal{S}$ could be combined with 
charge conjugation \emph{applied to the dark sector quarks}, 
\begin{eqnarray}
\mathcal{C}_{\rm dark}: \quad \mathcal{C}^{-1} \left\lbrace 
\begin{matrix}
    \mathfrak{q} \\ \mathfrak{r}
\end{matrix} 
\right\rbrace 
 \mathcal{C} & \longrightarrow & \left\lbrace 
\begin{matrix}
    \mathfrak{r} \\ \mathfrak{q}
\end{matrix} 
\right\rbrace  \, ,
\end{eqnarray}
where we emphasize that this operation swaps dark quarks in the
fundamental representation of SU($N_c$) with the opposite charge
dark quark partners in the anti-fundamental
representation.  When $\mathcal{S} \otimes \mathcal{C}_{\rm dark}$ is applied
to the Lagrangian of Eq.~(\ref{eq:L}), it yields
\begin{eqnarray}
  \parity:~i \mathfrak{q}^\dagger \bar{\sigma}^\mu D_\mu \mathfrak{q} + i \mathfrak{r}^\dagger \bar{\sigma}^\mu D_\mu \mathfrak{r}
  -\overline{m}_0 \left( \mathfrak{q} \mathfrak{r} + h.c. \right) 
  & \longrightarrow & 
  i \mathfrak{r}^\dagger \bar{\sigma}^\mu D_\mu \mathfrak{r} + i \mathfrak{q}^\dagger \bar{\sigma}^\mu D_\mu \mathfrak{q}
  -\overline{m}_0 \left( \mathfrak{q} \mathfrak{r} + h.c. \right), \nonumber \\
\end{eqnarray}
leaving the Lagrangian invariant.  For this parity to hold, it is obviously essential that dark quarks transform in vector-like representations under the gauge symmetries. Since this
symmetry swaps dark quarks with dark anti-quarks, 
Ref.~\cite{Bai:2010qg} 
found that this has important implications for the mesons of the
theory (see also Refs.~\cite{Berlin:2018tvf,Bernreuther:2019pfb,Abe:2024mwa}). The combined operations of $\mathcal{S}$ and $\mathcal{C}_{\rm dark}$
is closely analogous to G-parity that acts
on pions of the SM\@.  For dark pions, 
$\mathcal{G} \equiv \mathcal{S} \times \mathcal{C}_{\rm dark}$ takes $\pi^{(JM)} \stackrel{\mathcal{G}}{\longrightarrow} (-1)^J \pi^{(JM)}$. 
In Ref.~\cite{Bai:2010qg} it was emphasized that this symmetry guarantees the stability of the lightest dark meson that is odd under the dark G-parity - at least at the renormalizable level. 

On the other hand, when acting on baryons, the G-parity of Ref.~\cite{Bai:2010qg}, which includes charge conjugation of dark quarks, swaps baryons for anti-baryons. 
As a result, while G-parity may lead to relationships between the properties of baryons and anti-baryons, it does not act solely on baryons (or anti-baryons), and is thus completely distinct from our $\parity$-parity.

\section{Implications of $\parity$-Parity for Dark Matter Direct Detection}
\label{sec:GPandDD}

In the class of strongly coupled theories that respect $\parity$-parity, our results thus far demonstrate that there is a suite of higher dimensional operators involving electrically neutral dark baryons that are forbidden.  If dark matter is in fact composed of one of these neutral baryons, \textit{i.e.} the lightest dark baryon is electrically neutral and there is a viable mechanism to generate the cosmological abundance, then we can consider the implications of $\parity$-parity for dark matter direct detection.  The purpose of this section is to broadly identify the 
$\parity$-even and $\parity$-odd scattering processes of dark baryons off nuclei, emphasizing the weak constraints on $\parity$-even dark sector theories.  
We outline the contributions to the cross section rates using the past literature and provide leading estimates of the bounds.  Then in Section~\ref{sec:benchmark}, we will be more precise with a specific theory.

First, we remind the reader that elastic scattering cross sections are large in the presence of dimension-5 operators that are odd under $\parity$-parity.  In generic strongly coupled theories (without $\parity$-parity), an order one magnetic dipole moment is expected, \textit{i.e.}, $\mu_\chi \simeq \mathcal{O}(1) \times e/(2 m_\chi)$.
The scattering cross section off of nuclei through single photon exchange scales as $\sigma \propto \mu^2_\chi$, suppressed by (only) two powers of the dark matter mass.  
The corresponding Feynman diagram is shown in Table~\ref{tab:summary_bounds}.
Using the latest LZ constraints on direct detection of dark matter \cite{LZ:2022lsv,LZ2024}, this leads to a bound on the dark matter mass of 
$m_\chi \gtrsim 100 \; {\rm TeV}$ \cite{Eby:2023wem}. 
If the strongly coupled theory also violated CP, with an order one coefficient for an electric dipole moment, the bound is estimated to be much stronger, $\mathcal{O}$($10^4$)~TeV\@ (scaling the estimate from Ref.~\cite{Antipin:2014qva}, Eq.~(33), to the current LZ constraint).  
In theories that respect $\parity$-parity, all of these dimension-5 (as well as dimension-6) electromagnetic moments are forbidden, and so these bounds do not apply, significantly weakening the constraints from direct detection.  (For other prior estimates based on dimensional analysis, see Refs.~\cite{Cline:2013zca,Antipin:2014qva,Mitridate:2017oky,Cline:2021itd}.)

We now turn to the scattering processes that are allowed by $\parity$-parity.  First we consider the leading electromagnetic moment that is electromagnetic polarizability, \textit{e.g.}, see Refs.~\cite{Kribs:2009fy,Weiner:2012cb,Frandsen:2012db,Ovanesyan:2014fha,Appelquist:2015zfa,Kavanagh:2018xeh} for previous studies. 
This operator includes a two-photon exchange diagram shown in Table~\ref{tab:summary_bounds} that leads to  scattering off nuclei.  
The operator coefficient can be determined using dimensional analysis, following Refs.~\cite{Frandsen:2012db,Kavanagh:2018xeh}, 
or using lattice simulations in specific theories \cite{Appelquist:2015zfa}, or by ``scaling up'' the experimental value for the neutron. 
Even with this coefficient, we still need the nuclear structure factor for two photon exchange, but unfortunately this also has substantial nuclear physics uncertainties 
\cite{Ovanesyan:2014fha}. 
If we take the dark baryon polarizability estimate from the lattice calculation that was specific to Stealth 
Dark Matter ($N_c = 4$ with two heavier dark fermion flavors with charge $Q = \pm 1/2$) \cite{Appelquist:2015zfa}, we find the current constraint from LZ is approximately $m_\chi \gtrsim 300$~GeV\@.  If instead we simply ``scale up'' the neutron polarizability, we obtain approximately $m_\chi \gtrsim 180$~GeV, since the neutron has a somewhat smaller polarizability in comparion with the scalar dark baryon of Stealth Dark Matter.  Both of these estimates have at least a factor of $2$ uncertainty in the bound on the dark matter mass due to the uncertainties in the nuclear structure factor (that was shown as an uncertainty band for the result in Ref.~\cite{Appelquist:2015zfa}), and of course the precise value of the polarizability for a specific dark baryon will depend on $N_c$, $N_f$, and the dark fermion mass spectrum.  
Nevertheless, since the scattering cross section through the dark baryon polarizability scales as $\sigma \propto 1/m_\chi^6$, improvements in the experimental constraints from direct detection provide only limited ability to improve on these bounds.

\begin{table}[t]
    \centering
    \resizebox{\columnwidth}{!}{
    \begin{tabular}{|c|c|c|c|}
    \hline
     \thead{$\parity$ \\ Transformation} & Diagram & Operators & \thead{Direct Detection Constraints \\ (Naive Estimates)} \\
     \hline
     \Large{$\parity$-odd} & \raisebox{-0.5\totalheight}{\resizebox{0.25\columnwidth}{!}{\includegraphics{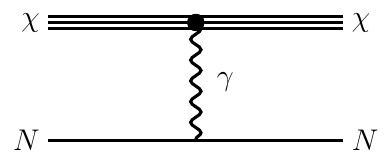}}} & \makecell{magnetic dipole moment \\ electric dipole moment \\ charge radius \\ anapole moment} & \makecell{(\textit{e.g.} magnetic dipole moment) \\ $m_\chi \gtrsim 100 $~TeV} \\
     \hline
     \multirow{3}{*}{\Large{$\parity$-even}} & \raisebox{-0.5\totalheight}{\resizebox{0.25\columnwidth}{!}{\includegraphics{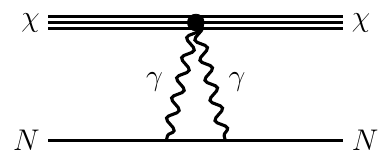}}} & \makecell{polarizability} & \makecell{$m_\chi \gtrsim \mathcal{O}(200)$~GeV}  \\
     \cline{2-4}
     & \makecell{\raisebox{-0.5\totalheight}{\resizebox{0.25\columnwidth}{!}{\includegraphics{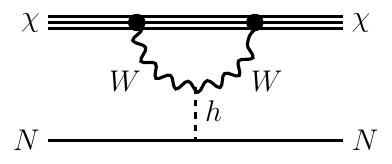}}}\\\resizebox{0.25\columnwidth}{!}{\raisebox{-0.5\totalheight}{\includegraphics{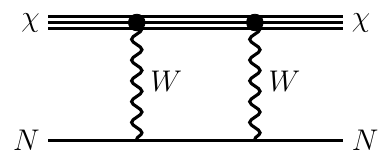}}}} & \makecell{electroweak \\ loop-induced} & \makecell{(\textit{i.e.} DM in SU(2)$_L$ \\ Triplet) \\ $m_\chi \gtrsim 200 $~GeV} \\
     \hline
    \end{tabular}
    }
    \caption{Diagrams giving rise to interactions between our DM and SM in direct detection experiments, operators they correspond to (see Section~\ref{subsec:GPinUV}), and the estimated lower bound they give rise to on DM mass assuming their Wilson coefficients are given by naive dimensional analysis. (Bounds are derived from current LZ experiment results \cite{LZ2024} and their exact value should be determined for models with different $N_{c}$,$N_{f}$ separately; the reported number is the conservative lower bound on DM masses, considering various uncertainties in this calculation \cite{Chen:2023bwg}.) 
    Our $\parity$-parity implies the \textbf{first row} diagrams are suppressed as the corresponding electromagnetic moments from Eqs.~\eqref{eq:PfIRMDM}--\eqref{eq:PfIRCRboson} are odd under $\parity$-parity, thus removing their constraints on DM mass.
    The \textbf{second row} pertains to the polarizability operators that are not forbidden by the $\parity$-parity. 
    The \textbf{third and fourth} rows correspond to electroweak loop-induced effects that generate non-electromagnetic operators of Eq.~\eqref{eq:EWloop}, see Ref.~\cite{Hill:2014yka} for their definitions. See the text for further details about the bounds on DM masses from the $\parity$-even operators.}
    \label{tab:summary_bounds}
\end{table}

Finally let us turn to the electroweak loop-induced diagrams.
These diagrams are shown in Table~\ref{tab:summary_bounds}, and have been carefully calculated in Refs.~\cite{Essig:2007az,Hisano:2010fy,Hill:2014yka,Hill:2014yxa,Chen:2018uqz,Bottaro:2021snn,Bottaro:2022one,Chen:2023bwg,Bloch:2024suj}. 
The scattering rate, and thus the experimental constraint, arising from these processes depends on the spin of DM and the SU(2)$_L$ representation in which the neutral baryon is embedded. The cross section as a function of DM mass has been computed in Refs.~\cite{Chen:2018uqz,Chen:2023bwg} for several representations and with different DM spins. 
In their calculation they used an expansion in $v_h/m_\chi$ which implies increasing uncertainties in the calculation as we consider DM masses below a TeV\@. 
For instance, if DM is the neutral component of a triplet (5-plet) of SU(2)$_L$ with spin half, the latest LZ bounds \cite{LZ2024} put a lower bound on the DM mass between approximately $0.2$ to $0.6$ ($1$ to $4$) TeV \cite{Chen:2023bwg}. 
While there will be some progress in probing lower  scattering cross sections in upcoming updates of direct detection searches, the large uncertainties in this calculation muddle a proper interpretation of these bounds on the DM mass, motivating further work to nail down the direct detection rates of the neutral components of (composite) electroweak multiplets that may be much lighter than the standard thermal estimates imply \cite{Mitridate:2017izz,Bottaro:2021snn}.

On the other hand, if the lightest baryon, that is made of constituent charged quarks, is a \textit{singlet} of SU(2)$_L$, the electroweak loop-induced processes are absent. This leaves the polarizability operators as the only source of direct detection signals in such theories, whose scattering rate drops rapidly with an increase in the DM mass as we discussed above. 
As a result, in these scenarios we expect the direct detection bounds will not be able to probe DM masses above a few hundred GeV in the foreseeable future. 
Instead, theories with composite electroweak-singlet DM are a well-motivated target for future collider searches. 
In a follow-up work, we will show that such singlet DM models arise naturally in theories with $\parity$-parity for many different values of $N_{c}$, $N_{f}$.

\section{The Minimal Fermionic Model with $\parity$-Parity} 
\label{sec:benchmark}

In this section we consider a specific model with $N_{c}=N_{f} = 3$. This is the most minimal $\parity$-preserving model (lowest values of $N_{c}$ and $N_{f}$) that yield neutral fermionic dark baryons. 
In the heavy quark limit, \textit{i.e.} $\Lambda_d/m_\chi \ll 1$, 
we can use the non-relativistic quark model \cite{DeRujula:1975qlm, Manohar:1983md, Georgi:1984zwz} to calculate the magnetic dipole moment and charge radius of the dark baryons. 
The results will be consistent with the operator transformation properties identified using $\parity$-parity from Section~\ref{sec:main}. 
Assuming one (of the two) neutral baryons is dark matter, we identify the viable range of dark baryon masses in this model and highlight some open mass ranges that were previously presumed to be ruled out.
Nonetheless, a concrete study of the viable parameter space rests upon a better understanding of the mass spectrum, especially the electroweak potential contribution, that we will address in a follow-up work. 
Like in Section~\ref{sec:GPandDD}, we do not specify the mechanism determining the abundance of dark matter and instead assume it arises from one of several possibilities.

\subsection{Baryons in the Minimal Fermionic Model}

Let us start by studying the spin-1/2 baryons in the spectrum and their transformation under the 
SU(2)$_L$ group. 
When $N_{c}=N_{f}=3$, the spectrum of spin-1/2 baryons is completely analogous to SM with three flavors of quarks,
with an octet of states that have the correct
spin-flavor wavefunctions to be baryons.
In the electroweak symmetric phase ($\Lambda_d > v_h$), 
the spin-1/2 octet of states split into a 
3-plet and 5-plet of SU(2)$_L$. 
As we emphasized in Section~\ref{sec:Lambdaabovev},
the neutral states $B_0$ do not have
magnetic moments or charge radii since the 
operators with $F_{W,\mu\nu}^3$ vanish and operators
with $B_{\mu\nu}$ are not generated in the dark sector.

In the electroweak broken phase ($\Lambda_d < v_h$), 
we can use the non-relativistic quark model, which is accurate in the very heavy dark quark limit \cite{Georgi:1984zwz}, to study properties of the model. 
In the quark model, a hadron is represented by a spin-flavor wavefunction consisting of its constituent valence quarks. Symmetry properties of the hadron suffice to uniquely determine the wavefunctions and some of its properties such as mass and electromagnetic moments; see Refs.~\cite{Georgi:1984zwz,Workman:2022ynf} for reviews 
of the quark model. 

Given the similarities between a dark sector with $N_{c}=N_{f}=3$ and the SM with three flavors, 
we use (or abuse) the SM notation to denote the 
spin-1/2 dark baryons spin-flavor wavefunctions as
\begin{eqnarray}
\label{eq:QMwavefunctions8}
| p \rangle & = & \frac{1}{\sqrt{18}} \left[ 	\left(  2 \mathfrak{q}_1\uparrow \mathfrak{q}_1\uparrow \mathfrak{q}_2\downarrow - \mathfrak{q}_1\uparrow \mathfrak{q}_1\downarrow \mathfrak{q}_2\uparrow - \mathfrak{q}_1\downarrow \mathfrak{q}_1\uparrow \mathfrak{q}_2\uparrow		\right) + perm.	\right], \\
| n \rangle & = & \frac{1}{\sqrt{18}} \left[ 	\left(  2 \mathfrak{q}_2\uparrow \mathfrak{q}_2\uparrow \mathfrak{q}_1\downarrow - \mathfrak{q}_2\uparrow \mathfrak{q}_2\downarrow \mathfrak{q}_1\uparrow - \mathfrak{q}_2\downarrow \mathfrak{q}_2\uparrow \mathfrak{q}_1\uparrow		\right) + perm.	\right], \nonumber \\
|  \Sigma^+ \rangle & = & \frac{1}{\sqrt{18}} \left[ 	\left(  2 \mathfrak{q}_1\uparrow \mathfrak{q}_1\uparrow \mathfrak{q}_3\downarrow - \mathfrak{q}_1\uparrow \mathfrak{q}_1\downarrow \mathfrak{q}_3\uparrow - \mathfrak{q}_1\downarrow \mathfrak{q}_1\uparrow \mathfrak{q}_3\uparrow		\right) + perm.	\right], \nonumber \\
| \Sigma^- \rangle & = & \frac{-1}{\sqrt{18}} \left[ 	\left(  2 \mathfrak{q}_2\uparrow \mathfrak{q}_2\uparrow \mathfrak{q}_3\downarrow - \mathfrak{q}_2\uparrow \mathfrak{q}_2\downarrow \mathfrak{q}_3\uparrow - \mathfrak{q}_2\downarrow \mathfrak{q}_2\uparrow \mathfrak{q}_3\uparrow		\right) + perm.	\right] ,\nonumber \\
| \Xi^0 \rangle & = & \frac{1}{\sqrt{18}} \left[ 	\left(  2 \mathfrak{q}_3\uparrow \mathfrak{q}_3\uparrow \mathfrak{q}_1\downarrow - \mathfrak{q}_3\uparrow \mathfrak{q}_3\downarrow \mathfrak{q}_1\uparrow - \mathfrak{q}_3\downarrow \mathfrak{q}_3\uparrow \mathfrak{q}_1\uparrow		\right) + perm.	\right] \nonumber ,\\
| \Xi^- \rangle & = & \frac{1}{\sqrt{18}} \left[ 	\left(  2 \mathfrak{q}_3\uparrow \mathfrak{q}_3\uparrow \mathfrak{q}_2\downarrow - \mathfrak{q}_3\uparrow \mathfrak{q}_3\downarrow \mathfrak{q}_2\uparrow - \mathfrak{q}_3\downarrow \mathfrak{q}_3\uparrow \mathfrak{q}_2\uparrow		\right) + perm.	\right] \nonumber, \\
| \Sigma^0 \rangle & = & \frac{1}{6} \left[ 	\left(  2 \mathfrak{q}_1\uparrow \mathfrak{q}_2\uparrow \mathfrak{q}_3\downarrow - \mathfrak{q}_1\uparrow \mathfrak{q}_2\downarrow \mathfrak{q}_3\uparrow - \mathfrak{q}_1\downarrow \mathfrak{q}_2\uparrow \mathfrak{q}_3\uparrow		\right) + perm.	\right] \nonumber, \\
| \Lambda \rangle & = & \frac{1}{\sqrt{12}} \left[ 	\left( \mathfrak{q}_1\uparrow \mathfrak{q}_2\downarrow \mathfrak{q}_3\uparrow - \mathfrak{q}_1\downarrow \mathfrak{q}_2\uparrow \mathfrak{q}_3\uparrow		\right) + perm.	\right], \nonumber 
\end{eqnarray}
where $\mathfrak{q}_{1,2,3}$ refers to the three flavors of dark quarks, $\uparrow / \downarrow$ denote the spin of each quark along $\hat{z}$, and we have used the standard terminology of SM baryons to refer to different states with the replacement $(u,d,s)\rightarrow (\mathfrak{q}_1,\mathfrak{q}_2,\mathfrak{q}_3)$.  We emphasize that our dark quark electric charges are different than SM quarks, and so the superscripts on baryons do not necessarily stand for their electric charges.  The notation ``$perm.$'' refers to all permutations of quarks at different positions in the wavefunction. 
While these wavefunctions are orthogonal, at this stage it is not clear if they are mass eigenstates. 

We have not yet assigned the electric charges ($\pm 1$ and $0$) to the three quark flavors $\mathfrak{q}_i$. 
We are free to assign the charges ($\pm 1$ and $0$) to these states in any order (and we have verified that properties of mass eigenstates are independent of this assignment). 
For the rest of this section we assign charges  $(Q_1,Q_2,Q_3)=(+1,-1,0)$, where $Q_i$ denotes the electric charge of the dark quark $\mathfrak{q}_i$.
With this charge assignment, the dark baryons have electric charges
\begin{eqnarray}
    \label{eq:Qbaryonslist}
    Q_{\Sigma^+} = 2 , ~~~~
    Q_{p,\Xi^0} = 1 , ~~~~
    Q_{\Sigma^0,\Lambda} = 0 , ~~~~
    Q_{n,\Xi^-} = -1 , ~~~~
    Q_{\Sigma^-} = -2 . 
\end{eqnarray}

In the electroweak symmetric phase, the eight baryons form
a 3-plet and a 5-plet of SU(2)$_L$. 
We can explicitly check this by studying the action of $J^\pm = (J^1 \pm i J^2)/\sqrt{2}$, with $J^i$ denoting different generators of SU(2)$_L$. Having done this, we find
\begin{equation}
    | \mbox{3-plet} \rangle \, = \, 
    \left(          \begin{matrix}
        - \cfrac{| p \rangle + | \Xi^0 \rangle }{\sqrt{2}} \\ 
        | \Sigma^0 \rangle \\
        - \cfrac{| n \rangle + | \Xi^- \rangle }{\sqrt{2}}
    \end{matrix}          \right) , \quad
     | \mbox{5-plet} \rangle \, = \, 
    \left(          \begin{matrix}
        | \Sigma^+ \rangle \\ 
         \cfrac{| p \rangle - | \Xi^0 \rangle }{\sqrt{2}} \\ 
        | \Lambda \rangle \\
        - \cfrac{| n \rangle - | \Xi^- \rangle }{\sqrt{2}} \\
        | \Sigma^- \rangle
    \end{matrix}          \right).
    \label{eq:3plet5plet}
\end{equation}
In the electroweak symmetric phase, these states are the mass eigenstates, while in the broken phase ($\Lambda_d < v_h$), since the flavor symmetry is also spontaneously broken, baryons of the same electric charge from these two electroweak multiplets can mix. 

We can check the action of $\parity$ on baryon states using their wavefunctions in Eq.~\eqref{eq:QMwavefunctions8} and the action of $\parity$ on quarks from Eq.~\eqref{eq:exampleG3}. We find
\begin{equation}
    \parity \, | \mbox{3-plet} \rangle \, = \, 
    \left( \begin{matrix}
0 & 0 & 1 \\
0 & -1 & 0 \\
1 & 0 & 0 \\
\end{matrix}
\right) | \mbox{3-plet} \rangle , \quad
     \parity \, | \mbox{5-plet} \rangle \, = \, 
    \left( \begin{matrix}
0 & 0 & 0 & 0 & 1 \\
0 & 0 & 0 & -1 & 0 \\
0 & 0 & 1 & 0 & 0 \\
0 & -1 & 0 & 0 & 0 \\
1 & 0 & 0 & 0 & 0 
\end{matrix}
\right) | \mbox{5-plet} \rangle ,
    \label{eq:Gaction3plet5plet}
\end{equation}
that agrees with results in Eqs.~\eqref{eq:exampleG3} and \eqref{eq:exampleG5} for the action of $\parity$ on multiplets of SU(2)$_L$. Note that the two electrically neutral baryons, $\Lambda$ and $\Sigma^0$, have different spin wavefunctions, which forbids their mixing and, thus, implies they will always be mass eigenstates. 

\subsection{Electromagnetic moments of the baryons}
\label{subsec:explicit_moments}

To check the predictions of $\parity$-parity, we can use the wavefunctions in Eq.~\eqref{eq:QMwavefunctions8} to explicitly calculate the magnetic dipole moment and the charge radius of neutral baryons of the minimal model. 
Our explicit calculation verifies the prediction of the $\parity$-parity for these observables, \textit{i.e.} they both vanish for neutral baryons $\Sigma^0$ and $\Lambda$.

In the quark model framework, the magnetic dipole moment of a hadron is given by the sum of moments of its constituent quarks. 
For the neutral baryons $\Sigma^0$ and $\Lambda$, through explicit calculation we find
\begin{equation}
\hat{\mu}_{\Sigma^0 ,\Lambda} = \frac{e}{2\overline{m}_{0}} \left(\begin{matrix}
0& -\frac{2}{\sqrt{3}} \\
-\frac{2}{\sqrt{3}}  & 0
\end{matrix}
\right) \, .
\label{eq:MDMnum}
\end{equation}
This result is in complete agreement with the predictions of $\parity$-parity, described in Section~\ref{subsec:GPinUV}:  the diagonal entries vanish, 
because they are odd under $\parity$-parity, while the off-diagonal entries can be non-zero, 
since $\Lambda$ and $\Sigma^0$ transform differently under $\parity$-parity.  Hence, the magnetic dipole transition moment $\langle \Sigma^0 | \sigma^{\mu\nu} F_{\mu\nu} | \Lambda\rangle$ is even under $\parity$-parity and \emph{allowed}.
Our result also agrees with a separate calculation of magnetic dipole moments for baryonic dark matter in Ref.~\cite{Aranda:2015jis}, see also Ref.~\cite{Antipin:2015xia} that provided a different argument for 
why the diagonal entries vanish based on an analogy with QCD.

It is worth emphasizing the highly nontrivial result that
the lightest neutral baryons have vanishing magnetic moments and yet, the \emph{magnetic transition moment} of the neutral baryons is allowed. 
As mentioned before, since $\Lambda$ and $\Sigma^0$ have different spin wavefunctions, they will not mix with each other in the IR, thus each of them is a mass eigenstate.
The magnetic transition moment provides a fascinating example of inelastic dark matter \cite{Tucker-Smith:2001myb,Chang:2008gd,Alves:2009nf,SpierMoreiraAlves:2010err,Chang:2010en,Kumar:2011iy}. 
If the mass splitting between the neutral baryon states 
were small, $\lesssim \mbox{few} \;100$~keV, 
the magnetic transition moment
could lead to nuclear (upscatter) recoil signal \cite{Chang:2010en,Kumar:2011iy,Bramante:2016rdh}
in direct detection experiments, or could lead to even more 
exotic signals such as luminous dark matter \cite{Feldstein:2010su}
with a sidereal daily modulating photon
signal in large detectors \cite{Eby:2019mgs,Eby:2023wem,Graham:2024syw}.
This underscores the importance of properly calculating the mass spectrum. 
In a follow-up work we will develop a variation method to calculate the mass of these states.  Our preliminary estimates suggest they  are not likely to be close enough in mass to generate an inelastic dark matter signal, at least in the absence of other modifications
to the model.

It is also worth studying the magnetic dipole moment of charged baryons in this model and predictions that $\parity$-parity implies for them. We can straightforwardly show that
\begin{equation}
\parity \, |p \rangle = |n\rangle,~ ~~\parity \, |n \rangle = | p \rangle,~ ~~\parity \, | \Xi^0 \rangle = | \Xi^- \rangle,~~~\parity \, | \Xi^- \rangle = | \Xi^0 \rangle,
    \label{eq:Goncharged1}
\end{equation}
\begin{equation}
\parity \, |\Sigma^+ \rangle = |\Sigma^-\rangle,~~~ \parity \, |\Sigma^- \rangle = |\Sigma^+\rangle.
    \label{eq:Goncharged2}
\end{equation}
When combined with the transformation of the photon field strength under $\parity$-parity, see Eq.~\eqref{eq:exampleGFmunu}, this suggests
\begin{equation}
    \mu_p = - \mu_n,~~~ \mu_{\Xi^0}=-\mu_{\Xi^-},~ ~~\mu_{\Sigma^+}=-\mu_{\Sigma^-},
    \label{eq:MDMchargedexplicitly}
\end{equation}
for the magnetic moments of the charged particles, while their transition magnetic moments obey
\begin{equation}
    \mu_{p\Xi^0} = - \mu_{n \Xi^-},~~~ \mu_{\Xi^0 p}=-\mu_{\Xi^- n}.
    \label{eq:transitionMDMchargedexplicitly}
\end{equation}

We can explicitly check these relations.
Doing the same calculation as for Eq.~\eqref{eq:MDMnum} for the charge baryon wavefunctions, we find 
\begin{equation}
\mu_{p,\Xi^0} = - \mu_{n,\Xi^-} = \frac{e}{2\overline{m}_{0}} \left(\begin{matrix}
\frac{5}{3}& 0 \\
0  & \frac{1}{3}
\end{matrix}
\right),
\label{eq:MDMnumcharged}
\end{equation}
\begin{equation}
\mu_{\Sigma^+} = - \mu_{\Sigma^-} = \frac{2e}{3 \overline{m}_{0}},
    \label{eq:MDMch2bench}
\end{equation}
in agreement with predictions of $\parity$-parity in Eqs.~\eqref{eq:MDMchargedexplicitly}-\eqref{eq:transitionMDMchargedexplicitly}.

We can similarly calculate other electromagnetic moments of neutral baryons. 
In particular, building on the general parametrization method of Refs.~\cite{PhysRevD.40.2997,PhysRevD.41.2865,PhysRevD.53.3754} in a large-$N$ expansion, Ref.~\cite{Buchmann:2000wf} developed a general expression for the charge radius of hadrons in the quark model. 
Using the existing symmetries of the theory, they are able to provide the charge radius, in the most general form, as a polynomial function of operators acting on the spin-flavor wavefunction - with the coefficients being functions of the spatial distribution of quarks in the baryon.
We explicitly checked that using their operator expansion for the charge radius operator (see Eq.~(3.6) in Ref.~\cite{Buchmann:2000wf}), we find the charge radius vanishes for the neutral baryons $\Sigma^0$ and $\Lambda$, while their transition charge radius operator can be non-zero,\footnote{Calculating the exact quantitative value of this transition moment requires an expression for the spatial distribution of quarks in the hadrons, which is beyond the scope of this work.} in agreement with predictions of the $\parity$-parity arguments.

\subsection{Phenomenology}
\label{subsec:pheno}

In this section we investigate the parameter space of the minimal fermionic DM model.  
The upshot of this section is that in the minimal fermionic DM model with $N_{c}=N_{f}=3$, a much larger range of DM masses become viable - as predicted by the existence of the $\parity$-parity.  
We will also see that the electroweak potential should be included for a complete study of the phenomenology in this mass range, while here we simply assume the lightest baryon is neutral and is not degenerate in mass with other baryons.

Before we study the bounds on the model, let us quickly comment on the baryon mass spectrum and the identity of the lightest dark baryon. 
Even though the quark model can be used to calculate the mass spectrum of this model \cite{DeRujula:1975qlm}, electroweak interactions were mostly neglected in previous studies. 
While the loop-level corrections have been worked out in the literature and can be included (amounting to a small mass splitting between particles in the same electroweak multiplet \cite{Cheng:1998hc,Gherghetta:1999sw,Feng:1999fu,Cirelli:2005uq}), the tree-level electroweak potential from the exchange of $W$ or $Z$ bosons has not been studied in detail.

To check the relevance of these corrections, we should compare the range of these forces to the size of dark baryons
\begin{equation}
r_0 \approx \frac{1}{m_\chi \alpha_\chi (r_0)},
    \label{eq:r0}
\end{equation}
where $m_\chi$ is DM mass, $\alpha_\chi (r_0)$ is the dark fine structure constant evaluated at the Bohr radius of the baryon, $r_0$. 
When $r_0 m_W \gg 1$, as is the case in the SM, the electroweak potential is exponentially suppressed on the scale of the dark baryon. Consequently, we can use the quark model calculation of Ref.~\cite{DeRujula:1975qlm} with only the photon and gluon potentials. 
In the opposite limit of $r_0 m_W \ll 1$, the electroweak symmetry is restored on the scales of the dark baryons, and 
the eigenstates are nearly exact SU(2)$_L$ states from Eq.~\eqref{eq:3plet5plet}.
To obtain the mass spectrum of the dark baryons for intermediate values of $r_0 m_W$, the potential due to the exchange of $W$ and $Z$ boson exchanges should be included.

While we expect the neutral baryons $\Lambda$ and $\Sigma^0$ to be lighter than charged baryons, which is shown to be the case in the limit of $r_0 m_W \ll 1$ \cite{Cirelli:2005uq}, 
further studies are warranted to determine which one of them will be lighter, and thus is the DM candidate.
In the electroweak symmetric case, $r_0 m_W \ll 1$, the 5-plet mass is slightly larger than the 3-plet mass,
due to the larger group theory factors.
This is analogous to the electromagnetic corrections
to SM baryons \cite{Gasser:1982ap}. 
But given that we are considering dark baryon masses 
that can be close to the electroweak scale, 
we allow for either the neutral baryons $\Lambda$ or $\Sigma^0$ 
to be the lightest state and report the 
direct detection bounds in both cases.

\begin{figure}[t!]
\begin{center}
\resizebox{0.8\columnwidth}{!}{
\includegraphics[scale=1]{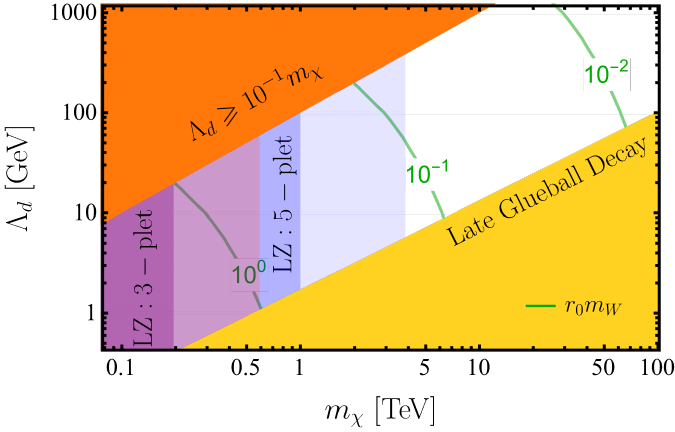}
}
\end{center}
\caption{Parameter space in the minimal fermionic DM model of $N_{c}=N_{f}=3$.  The yellow region shows the bounds from late glueball decay, i.e., requiring the lifetimes to be shorter than the onset of BBN\@.  The electroweak loop-induced interactions of dark matter that is part of an SU(2)$_L$ 3-plet (5-plet) can be probed in the purple (blue) region; 
the darker (lighter) shade region denotes the most (least) conservative bounds given uncertainties in the calculation as described in the text. 
In the orange region the quark masses are comparable to the confinement scale, hence the non-relativistic quark model predictions become unreliable (nonetheless, the $\parity$-predictions are still valid in this region too).
The green contours show $r_0 m_W$ that is close to unity across most of the viable DM mass range, highlighting the importance of including the neglected electroweak potential in the quark model for a proper study of the mass spectrum. 
If dark matter has an order one magnetic dipole moment, 
the \emph{entire white region} would be ruled out by 
bounds from direct detection.   
The upshot of our work is that $\parity$-parity removes the low dimensional electromagnetic form factors and thus opens up the white region of parameter space. }
\label{fig:splitting_degQ}
\end{figure}

In Figure~\ref{fig:splitting_degQ} we show the viable parameter space of this minimal model. 
The regions in blue and purple can be probed by direct detection mediated by 
electroweak loop-induced diagrams (see Table~\ref{tab:summary_bounds}) using current LZ results \cite{LZ:2022lsv}.\footnote{The scattering cross sections in Ref.~\cite{Chen:2023bwg} are calculated for when DM is part of a Majorana multiplet of SU(2)$_L$. In our theory, on the other hand, we are studying Dirac multiplets. However, we use the results of Ref.~\cite{Chen:2023bwg} as a proxy for the direct detection region that can be probed in our theory, leaving a more careful treatment for future work.} 
There are uncertainties in the calculation performed by Ref.~\cite{Chen:2023bwg}, in part because the calculation was performed doing an 
expansion in $v_h/m_\chi$ (leading to larger uncertainties as $m_\chi$ approaches $v_h$).
If we apply a broad brush to these uncertainties, 
we can use Ref.~\cite{Chen:2023bwg}
to estimate the current bounds \cite{LZ2024} for the neutral component of an SU(2)$_L$ 5-plet to be between $1$-$4$~TeV
(\textit{i.e.}, the $\Lambda$ baryon in our theory)
and for an SU(2)$_L$ triplet to be between $0.2$-$0.6$~TeV, 
(\textit{i.e.}, the $\Sigma^0$ baryon in our theory).
In this figure we also show the part of the parameter space 
ruled out by late decaying glueballs. 
Glueballs become long-lived when 
$\bar{m}_0 \gg \Lambda_d$, 
and if the lifetime exceeds approximately 
$\tau \gtrsim 0.1 ~$sec, 
their decays back into high energy SM particles 
can disrupt the elemental abundances resulting from Big Bang Nucleosynthesis (BBN) \cite{Jedamzik:2006xz,Kawasaki:2017bqm}. 
Glueball decay in the $\parity$-parity theories that we
consider proceeds through the dimension-8 operators
into SM gauge bosons.  
For the purposes of the figure, 
we use $m_{0^{++}} \simeq 7 \Lambda_d$ \cite{Morningstar:1999rf,Mathieu:2008me} 
and the calculations in \cite{Juknevich:2009ji,Juknevich:2009gg}
to delinate the region of $m_\chi$-$\Lambda_d$ space
that is viable.

Figure~\ref{fig:splitting_degQ} shows contours of $r_0 m_W$ as well. 
For large enough DM masses, $r_0 m_W \ll 1$, we expect the multiplets in Eq.~\eqref{eq:3plet5plet} to be mass eigenstates since this parameter region is in the electroweak restored phase.  
For most of the mass range shown in the figure, $r_0 m_W$ is close enough to unity that electroweak potential effects are anticipated to affect the mass spectrum of the baryonic states. 
In a follow-up work we provide further details about the mass spectrum in this model, as well as models with different values of $N_{c}$ and $N_{f}$.

, we find that $\parity$-parity suppresses 
the elastic scattering cross section for direct detection searches, giving rise to a viable DM mass range that, prior to our paper,   was presumed to be ruled out (the white region shown in Figure~\ref{fig:splitting_degQ}). 
This assumes dark matter is made exclusively of dark baryons, 
and dark mesons either annihilate or decay fast enough to evade astrophysical bounds. 
If there were cosmologically stable dark mesons, 
their lower dimensional moments (charge radius) are also absent.
If the dark mesons made up a sizable component of dark matter, they could also be visible in direct detection through the same $\parity$-parity even operators that were allowed for scalar dark baryons.

\section{Conclusion and Discussion}
\label{sec:conclusion}

In theories with a new dark confining gauge group SU($N_c$) and a 
set of dark fermions transforming as vector-like representation of the dark group and SU(2)$_L$ (with zero hypercharge), we find the electrically neutral dark baryons do not have dimension-5 or dimension-6 electromagnetic moments 
(electric and magnetic dipole moments, charge radius, and the anapole moment) due to symmetries that we have identified in the theory.  In the electroweak unbroken phase, the symmetry responsible for this is SU(2)$_L$ itself.  In the electroweak broken phase, we identified a new symmetry called $\parity$-parity that leads to the same result. 
If dark matter is one of the electrically neutral dark baryon states, the absence of low dimensional electromagnetic moments implies the cross section for elastic scattering off the SM
is highly suppressed.  Scattering through electroweak 
loop-induced contributions remains possible, as does 
dimension-7 polarizability.  
The suppression of $\parity$-parity symmetric dark baryon 
dark matter 
direct detection signal is an automatic
consequence of the underlying symmetries of the model.
Our results apply to all such vector-like confinement models 
with $\theta_\chi = 0$, where the $\parity$-parity remains unbroken in the confined phase of the theory.

Electroweak loop-induced elastic scattering is possible for neutral dark baryons in a (nontrivial) multiplet of SU(2)$_L$, and this process can be used to probe such a dark matter candidate in current and future direct detection experimental searches. 
On the other hand, if the confining description contains a dark baryon in a \emph{singlet} of SU(2)$_L$, then the leading interaction is just the electromagnetic polarizability. Given the rapid decline of the elastic scattering cross section through polarizability,  
once the mass of such a dark baryon is above a few hundred GeV, all its direct detection signals are completely suppressed.
Instead, colliders become one of the few ways to search for the signals. 
In a follow-up work we will show that for many different values of $N_{c}$ and $N_{f}$ there are dark baryons in SU(2)$_L$ singlets, highlighting the importance of future investigations of collider phenomenology of vector-like confinement with $\parity$-parity.

While $\parity$-parity forbids the electromagnetic moments of dark hadrons, it does not necessarily forbid the transition moments. 
Thus, if the spectrum includes at least two states, such as two neutral dark baryons that are close in mass, 
the model could naturally give rise to inelastic scattering-mediated signals in direct detection experiments and other astrophysical contexts. 
We demonstrated this possibility explicitly in the specific
case of $N_c = N_f = 3$, where the dark baryon spectrum
contains a dark $\Lambda$ and a dark $\Sigma^0$ that have no
magnetic moments, but they \emph{do} have a magnetic 
transition operator between the states.  
As a result, the question of whether the direct detection signal is suppressed depends sensitively on the mass spectrum of the model and should be addressed carefully. That said, for the range of confinement scales allowed by existing astrophysical bounds, we expect such approximate degeneracies most likely require significant tuning to render the inelastic scattering signal detectable in direct detection experiments. 

The implications of the symmetries identified in this paper, specifically $\parity$-parity, have wide applicability to theories with a dark confining sector:
\begin{itemize}

    \item They apply to models with any values of $N_f$ and $N_c$, as well as any ratio of the dark quark mass to the confinement scale.

    \item While we focused on dark quarks that transform as fundamentals of the dark confining SU($N_c$) group, our results apply to models with dark quarks in any SU($N_c$) representations, since $\parity$-parity and the dark group commute.
    
    \item While we assumed there was only one vector-like multiplet of dark quarks in the $N_f$ representation of SU(2)$_L$, our results can be extended to models with more multiplets, each transforming under different representations of SU(2)$_L$ (subject only to the restriction that the dark
    sector confines in the IR).

    \item Additional interactions could exist in the dark sector, with the only restriction that such interactions respect the vector-like nature of the dark quarks, leaving $\parity$-parity invariant.

\end{itemize}

Our study can be extended in many different directions. 
Foremost among them is a study of models with different values of $N_{c}$, $N_{f}$ and their phenomenology. 
As mentioned above, in an upcoming work we will discuss direct and indirect detection signatures of models with different values of $N_{c}$, $N_{f}$, while leaving a more detailed collider study for future work.
We also intend to study the mass spectrum of the models, since this not only impacts many of the signals of the theory, but is directly relevant to ensuring a viable dark matter candidate exists.
In this work we simply assumed the lightest dark baryon is electrically neutral, but this assumption is not guaranteed and needs to be checked for each specific model.

If kinematically accessible, the class of models discussed in this paper can have a large production rate at colliders, giving rise to a host of different signals. 
Depending on details of the model, \textit{e.g.}, number of colors and flavors or whether the dark quarks are heavy or light compared to the confinement scale, such models can give rise to a diverse class of prompt and/or long-lived particle signals at LHC and future colliders.
Vector-like confinement was long ago recognized to lead to signals at the LHC \cite{Kilic:2009mi}, such as production of dark quarks through Drell-Yan-like processes or through the mixing of SM vector bosons with the dark $\rho$ mesons, with earlier studies including Refs.~\cite{Kilic:2009mi,Kribs:2018oad,Kribs:2018ilo,Cheng:2021kjg}. 
These quarks can then shower and form dark mesons and baryons, giving rise to both detectable tracks and missing energy. 
Depending on their lifetimes, mesons and excited baryons can also give rise to unique long-lived particle signals, such as quirky tracks \cite{Kang:2008ea} or tumblers \cite{Dienes:2021cxr}. 
Similar signals can be searched for at a future Muon Collider. We will explore these signals in future work.

We did not study the relic abundance calculation of dark baryon dark matter. Where we make statements about the suppression of a direct detection signal, we simply assumed the dark matter abundance arose from an unspecified mechanism. 
While this may seem unsatisfactory, one of the features of confining dark sectors is the diversity of possible mechanisms and range of dark matter scales that could arise.  
Our contribution in this paper is to delineate the model space in which 
symmetries including $\parity$-parity can arise, and its implications for dimension-5 and dimension-6 electromagnetic
moments.  
Mapping this model space onto the various symmetric and asymmetric composite dark matter production mechanisms is left to future work.

Indirect detection through dark matter annihilation may also be an exceptionally important signal if the astrophysical abundance of dark baryons (as dark matter) contains a \emph{symmetric} amount of baryons and anti-baryons. 
The indirect detection signal strongly depends on the dark matter transformation under SU(2)$_L$, with some candidates already highly constrained by astrophysical observations  \cite{Cohen:2013ama,Asadi:2016ybp,Mitridate:2017izz,Baumgart:2017nsr,Baumgart:2018yed,Baumgart:2023pwn}. 
There are several important differences between dark baryon electroweak multiplets and the elementary candidates that have been the main focus in the past:  Dark baryons necessarily come in complex representations (\textit{e.g.}, a complex scalar or a Dirac fermion) given their dark baryon number transformation property, whereas the focus of many direct detection and indirect detection studies of elementary candidates have been on Majorana states.  
Additionally, dark baryons can annihilate to dark mesons, which could potentially dominate other annihilation channels \cite{Mitridate:2017oky}. 
Depending on the structure of the theory, some of the mesons in such models can be long-lived due to G-parity \cite{Bai:2010qg}, in which case this annihilation signal could be detectable in celestial-body focused searches as well \cite{Leane:2021ihh}.

Finally, as illustrated in Refs.~\cite{Kribs:2009fy,Buckley:2012ky}, closely-related parities exist in chiral confining dark sectors as well. It would be interesting to explore the most general symmetries, \textit{i.e.}, applicable to the broadest class of confining dark sectors possible.
In particular, investigating the case of $\theta_\chi \neq 0$ to see if the $\parity$-parity is broken (or remains unbroken), and whether the dimension-5 and dimension-6 electromagnetic moments remain suppressed.

\section*{Acknowledgments}

We are especially in debt to Yang Bai, Spencer Chang, and Riccardo Rattazzi for discussions about $\parity$-parity. We also thank Austin Batz, Elias Bernreuther, Tom Bouley, Kevin Langhoff, Ben Lillard, Hitoshi Murayama, Juri Smirnov, and Tien-Tien Yu for helpful discussions. 
The work of P.A. and G.D.K. is supported by the U.S. Department of Energy under grant number DE-SC0011640. 
P.A. thanks the Aspen Center for Physics, which is supported by National Science Foundation grant PHY-2210452, where part of this work was
completed.
G.D.K. thanks the Mainz Institute for Theoretical
Physics and the CERN Department of Theoretical Physics
where part of this work was completed. 
The work of C.J.H.M is supported by University of Oregon's Presidential Undergraduate Research Scholarship and Summer Undergraduate Research Fellowship.


\appendix

\section{$\parity$-Parity and the SM}
\label{appx:GPV}

\subsection{Absence of $\parity$-parity for SM baryons}

To begin this discussion, 
a natural question to first consider is precisely why there
is no analogue to $\parity$-parity for Standard Model baryons.
The existence of magnetic dipole moments for all
of the (electrically neutral) SM baryons provides clear experimental evidence against any such parity.  Yet, there are several similarities 
of the dark sector to the low energy 
effective theory of baryons suggesting that the SM
has \emph{almost} all of the
ingredients to have its own $\parity$-parity.
In the EFT below electroweak symmetry breaking
consisting of SU(3)$_c \times$~U(1)$_{\rm em}$, 
the quarks of the SM are vector-like.
The vector-like ``current'' masses for the quarks vary over
a large range (from well above $\Lambda_{\rm QCD}$
to well below this), but as is well known from the early days of the
quark model, baryon magnetic moments are well 
described by ``constituent'' quark masses that 
divide the baryon mass into approximately equal 
effective masses for each of the quarks.  
Considering just the three light(er) quarks
$u$, $d$, and $s$, 
the SM baryon magnetic moments can be
described as an approximate SU(3)$_f$ flavor 
symmetry \cite{Parreno:2016fwu}, where the
constituent quark masses are taken to be the same
in a leading  approximation.

At this point, the only distinction between the SM
and the minimal dark sector model with $N_c = N_f = 3$
described in Section~\ref{sec:benchmark} 
is the set of electric charges of the quarks
of the SM:  $u(+2/3), d(-1/3), s(-1/3)$ instead of 
$\mathfrak{q}_1(+1), \mathfrak{q}_2(-1), \mathfrak{q}_3(0)$ for the minimal dark sector
that preserves $\parity$-parity.  If indeed the quarks of the SM had electric charges the same as the minimal dark sector 
(and the constituent masses were consistent with SU(3)$_f$ symmetry), 
the ``real'' $\Lambda$ and $\Sigma^0$ of QCD would have 
had no magnetic dipole moments.

Of course there is no way to actually realize electric 
charges $(q,0,-q)$ among three light quarks in the SM
without changing \emph{both} the SU(2)$_L$ representation
and hypercharges of the quarks.
For instance, just removing hypercharge for the quarks, 
$Y_q \rightarrow 0$, 
results in a set of light and heavy quarks 
with electric charges
$(1/2,-1/2)$.  While this theory does have a 
well-defined $\parity$-parity, there are no electrically
neutral baryons for $N_c = 3$ and $Y_q =0$, and so $\parity$-parity would not be immediately relevant for forbidding electromagnetic moments.

\subsection{$\parity$-parity in the ``hyperchargeless'' SM}

The observation that $\parity$-parity would be present 
in the SM if hypercharge were absent, $Y_q \rightarrow 0$,
naturally leads to our next discussion of how the SM
violates $\parity$-parity, and how this is 
communicated to the dark sector.

We need to first remind the reader that $\parity$-parity
requires the SM electroweak gauge bosons transform under charge conjugation according
to Eqs.~\eqref{eq:Wchargeconjugation} and \eqref{eq:Bchargeconjugation}.
Matter that transforms under representations of SU(2)$_L$,
will also remain invariant under $\parity$-parity,
so long as there are no other interactions (in particular
hypercharge, as we will see explicitly below).

Consider first a simplified SM in which the hypercharges
of the quarks, leptons, and Higgs scalar doublet are 
taken to vanish
$Y_f = Y_H = 0$.
In this ``hyperchargeless'' SM, the action of $\parity$-parity on the 
left-handed quark doublets ($Q$), lepton doublets ($L$),
and Higgs doublet ($H$) is
\begin{eqnarray}
       Q & \longrightarrow & S Q \label{eq:Qshift}  \\
       L & \longrightarrow & S L \label{eq:Lshift} \\  
       H & \longrightarrow & S H \label{eq:Hshift} \, ,
\end{eqnarray}
where $S = \exp(i \pi J_2)$, and in this case, $S$ acts on the doublet representation of SU(2)$_L$.
These $S$ transformations ensure all (hyperchargeless) 
SM interactions
with the $W^i_\mu$ gauge bosons are invariant.
For the fermions this is evident from Eq.~\eqref{eq:Hsimtwocomponent} applied to the SM fermion doublets, while for the Higgs scalar doublet one simply has
to expand the covariant derivative (with $Y_H = 0$)
to see that it remains invariant.

In the hyperchargeless SM, the Yukawa interactions are 
also invariant, since they involve bilinear
SU(2)$_L$ contractions $H^\dagger Q$, $H^\dagger L$, and $H^T i \tau_2 Q$.
Under the shifts Eqs.~(\ref{eq:Qshift})-(\ref{eq:Hshift}) the first two result in $S^\dagger S = 1$,
while the third becomes $S^T i \tau_2 S = i \tau_2$,
and so all SU(2)$_L$ contractions within the Yukawa
terms of the (hyperchargeless) SM are invariant.

\subsection{$\parity$-parity violation in the full SM}

The presence of hypercharge violates 
$\parity$-parity, since under an $S$ rotation
of quark doublets (and lepton doublets) 
simultaneous with charge conjugation of 
the hypercharge gauge boson $B_\mu$, 
the hypercharge part of the kinetic term
\begin{eqnarray}
    \parity:  Q^\dagger (-i g' Y_q B_\mu) Q
    & \longrightarrow & Q^\dagger S^\dagger (-i g' Y_q B_\mu^c) S Q \\
    & = & Q^\dagger (+i g' Y_q B_\mu) Q \, ,
\end{eqnarray}
changes sign.  It is this sign-flip that shows $g'$ is a spurion for $\parity$-parity breaking.

Notice also that SU(2)$_L$-singlet fermions $f$ 
also undergo a sign-flip under $\parity$-parity,
\begin{eqnarray}
    \parity:  f^\dagger (-i g' Y_u B_\mu) f
    & \longrightarrow & f^\dagger (-i g' Y_q B_\mu^c) f \\
    & = & f^\dagger (+i g' Y_q B_\mu) f \, .
\end{eqnarray}
This is a necessary consequence following the requirement that ordinary charge conjugation of the photon, $A_\mu^c \rightarrow -A_\mu$, remains a symmetry of $U(1)_{\rm em}$. 
Moreover, it also means that, even though quark and lepton
interactions with the hypercharge gauge boson 
violate $\parity$-parity, the Yukawa interactions
of the full SM remain invariant.  This is simply a 
consequence of flipping the hypercharge assignment
of the quarks, leptons, and the Higgs doublet simultaneously.

Having concluded that it is \emph{only} hypercharge that violates $\parity$-parity, we see that there is a clear distinction between custodial SU(2) symmetry breaking and $\parity$-breaking:  while both are broken by hypercharge, custodial SU(2) is also broken by the difference between the top and bottom Yukawa couplings.  Hence, $\parity$-parity and custodial SU(2) symmetry are completely distinct.

\subsection{Implications of $\parity$-parity violation on the dark sector}

The presence of $\parity$-parity violation
in the SM interactions with hypercharge
implies there is a higher-loop SM contribution
to operators that are otherwise forbidden.
Consider first the case where $\Lambda_d > v_h$,
where we can write the higher dimensional 
operators involving the hypercharge field strength
for fermionic dark baryons
\begin{eqnarray}
\frac{c_{5,a}}{\Lambda_d} \;
\bar{\chi}_{B,a} \, \sigma^{\mu\nu} (\gamma^5) \, \chi_{B,a} \, F_{B,\mu\nu} \label{eq:hyperdim5fllv} \\
\frac{c_{6,a}}{\Lambda_d^2} \;
\bar{\chi}_{B,a} \, \gamma^\mu (\gamma^5)\,  \chi_{B,a} \, \partial^\nu F_{B,\mu\nu}
\end{eqnarray}
and scalar dark baryons 
\begin{equation}
\frac{c'_{6,a}}{\Lambda_d^2} \;
\phi_{B,a}^\dagger \, \overset{\text{\tiny{$\longleftrightarrow$}}}{\partial^{\mu}} \, \phi_{B,a} \, \partial^\nu  F_{B,\mu\nu} \, .
\label{eq:hyperdim6sllv}
\end{equation}
Under the action of $\parity$-parity, 
$(\bar{\chi}_{B,a} \chi_{B,a})$
and $(\phi^\dagger_{B,a} \phi_{B,a})$
are invariant but 
$F_{B,\mu\nu} \rightarrow (F_{B,\mu\nu})^c = -F_{B,\mu\nu}$.  This shows that if $\parity$-parity
were exact, these operators would be 
forbidden, thus there are no dimension-5 or
dimension-6 electromagnetic moments
for the dark baryons.

However, the breaking of $\parity$-parity
by hypercharge causes 
Eqs.~(\ref{eq:hyperdim5fllv})-(\ref{eq:hyperdim6sllv})
to be generated at loop-level through
interactions with the SM\@. 
We show some representative leading diagrams
in Figure~\ref{fig:GPV}:
\begin{figure}
    \centering
    \resizebox{0.9\columnwidth}{!}{
    \includegraphics{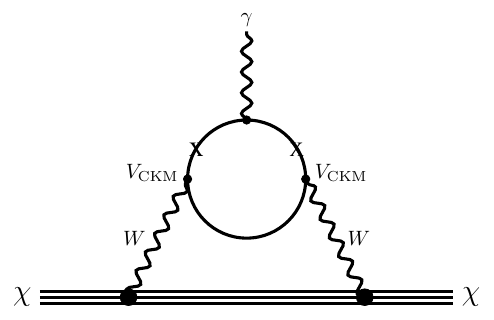}
    \hspace{0.2in}
    \includegraphics{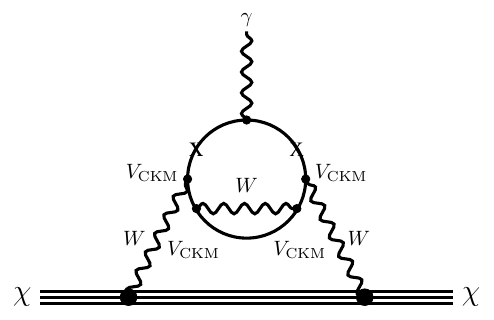}
    }
    \caption{Loop diagrams that could give rise to an explicit breaking of the $\parity$-parity and generating magnetic dipole moment or charge radius (\textbf{left}) and electric dipole or anapole moments (\textbf{right}), when DM is in a multiplet of SU(2)$_L$. In addition to violating $\parity$-parity, the SM also provides CP violation proportional to its Jarlskog invariant $J = \mathrm{Im} ~\mathrm{Tr}\left( V_\mathrm{CKM} V^\dagger_\mathrm{CKM} V_\mathrm{CKM} V^\dagger_\mathrm{CKM} \right)$; this is why the CP-odd operators (electric dipole and anapole moments) are only generated at three loops, while the CP-even ones are generated at two loops.}
    \label{fig:GPV}
\end{figure}
In the left panel we show the diagrams leading to CP-even operators, namely 
the dark baryon's magnetic dipole moment 
or charge radius, while
in the right panel we show the diagrams 
leading to CP-odd operators, namely 
the dark baryon's electric dipole moment 
and anapole moment. 

We will not attempt a rigorous estimate 
of these diagrams, nor even to demonstrate 
that they necessarily give a non-zero result.
However, we see no symmetry or other reason 
to suggest the diagrams vanish,
and so a two-loop contribution
to the magnetic dipole moment or the charge radius and a
three-loop contribution to the electric dipole moment 
or the anapole moment are expected.
In both cases, 
there is substantial suppression
from the weak couplings and loop factors
so as to make these contributions
completely subdominant to the electroweak 
loop-induced contribution
discussed in Section~\ref{sec:GPandDD}.

\section{Alternative $\parity$-parities}
\label{app:alternativeH}

In Section~\ref{sec:lambdabelowv}, we emphasized
that $\parity$-parity acts as charge conjugation
on the hypercharge gauge boson such that 
the photon retains its standard 
charge conjugation. In the electroweak 
unbroken phase, this led to the 
conclusion that the higher dimensional
operators involving the hypercharge
field strength, $F_{B,\mu\nu}$, 
given in Eqs.~(\ref{eq:hyperdim5fllv})-(\ref{eq:hyperdim6sllv}), 
were forbidden by $\parity$-parity.
But, since the hypercharge interactions
of SM fermions are themselves violating
$\parity$-parity, these operators can 
be regenerated through loops of SM fermions,
shown in Figure~\ref{fig:GPV}.

There is an alternative, that appears at least
consistent in the electroweak unbroken phase.
Consider $\parity'$-parity that has the
identical transformations for 
SU(2)$_L$ fields, but instead acts
trivially on $B_\mu$, i.e., 
\begin{eqnarray}
    \parity': \; B_\mu \rightarrow B_\mu \, .
\end{eqnarray}
$\parity'$-parity has the feature that the
SM Lagrangian is \emph{invariant}. 
Namely, the full set of quark, lepton, 
and Higgs multiplet interactions are
invariant under $\parity'$. 

Focusing on the interactions between the
dark sector and the SM, we see that this 
also means that the operators given above 
in Eqs.~(\ref{eq:hyperdim5fllv})-(\ref{eq:hyperdim6sllv}) are 
\emph{allowed} by $\parity'$-parity.
Now since the dark sector does not 
transform under hypercharge, 
they are not generated purely by the dark 
sector.  But, since the SM quarks and leptons
do transform under hypercharge, these operators
can be generated through loops of SM fields --
preicsely the same loops that we showed in
Figure~\ref{fig:GPV}.  Hence, in the electroweak
unbroken phase, it would appear
that there is no physical difference between 
$\parity$-parity and $\parity'$-parity:
operators involving the hypercharge
field strength, $F_{B,\mu\nu}$, 
are generated by the same SM loops 
in both cases.

One might be tempted to think that we have
uncovered a new exact symmetry of the SM, 
$\parity'$-parity.  However, this is not the
case.  The critical reason that dark sector
confinement preserved $\parity$-parity
(and $\parity'$-parity) is because it is
vector-like, and so the $\mathcal{S}$
rotation is performed on both the 
fundamental and anti-fundamental fields.
QCD confinement also generates a 
condensate between fundamental and 
anti-fundamental fields, but 
the $\mathcal{S}$ rotation only 
transforms the left-handed fields,
not the right-handed ones, due to the
chiral structure of the SM fermion content.
Hence, QCD confinement breaks 
$\parity'$-parity (and $\parity$-parity).
Whether this observation is or not useful
in other contexts remains an interesting 
open question.

\end{spacing}

\bibliography{ref}

\providecommand{\href}[2]{#2}\begingroup\raggedright\begin{thebibliography}{100}

\bibitem{Cirelli:2024ssz}
M.~Cirelli, A.~Strumia, and J.~Zupan, ``{Dark Matter},'' \href{https://arxiv.org/abs/2406.01705}{{\ttfamily arXiv:2406.01705 [hep-ph]}}.

\bibitem{Sakharov:1967dj}
A.~D. Sakharov, ``{Violation of CP Invariance, C asymmetry, and baryon asymmetry of the universe},'' \href{https://dx.doi.org/10.1070/PU1991v034n05ABEH002497}{{\em Pisma Zh. Eksp. Teor. Fiz.} {\bfseries 5} (1967) 32--35}.

\bibitem{Bai:2013xga}
Y.~Bai and P.~Schwaller, ``{Scale of dark QCD},'' \href{https://dx.doi.org/10.1103/PhysRevD.89.063522}{{\em Phys. Rev. D} {\bfseries 89} no.~6, (2014) 063522}, \href{https://arxiv.org/abs/1306.4676}{{\ttfamily arXiv:1306.4676 [hep-ph]}}.

\bibitem{Okun:1980mu}
L.~B. Okun, ``{THETA PARTICLES},'' \href{https://dx.doi.org/10.1016/0550-3213(80)90439-3}{{\em Nucl. Phys. B} {\bfseries 173} (1980) 1--12}.

\bibitem{Kribs:2009fy}
G.~D. Kribs, T.~S. Roy, J.~Terning, and K.~M. Zurek, ``{Quirky Composite Dark Matter},'' \href{https://dx.doi.org/10.1103/PhysRevD.81.095001}{{\em Phys. Rev. D} {\bfseries 81} (2010) 095001}, \href{https://arxiv.org/abs/0909.2034}{{\ttfamily arXiv:0909.2034 [hep-ph]}}.

\bibitem{Hambye:2009fg}
T.~Hambye and M.~H.~G. Tytgat, ``{Confined hidden vector dark matter},'' \href{https://dx.doi.org/10.1016/j.physletb.2009.11.050}{{\em Phys. Lett. B} {\bfseries 683} (2010) 39--41}, \href{https://arxiv.org/abs/0907.1007}{{\ttfamily arXiv:0907.1007 [hep-ph]}}.

\bibitem{Bai:2010qg}
Y.~Bai and R.~J. Hill, ``{Weakly Interacting Stable Pions},'' \href{https://dx.doi.org/10.1103/PhysRevD.82.111701}{{\em Phys. Rev. D} {\bfseries 82} (2010) 111701}, \href{https://arxiv.org/abs/1005.0008}{{\ttfamily arXiv:1005.0008 [hep-ph]}}.

\bibitem{Fok:2011yc}
R.~Fok and G.~D. Kribs, ``{Chiral Quirkonium Decays},'' \href{https://dx.doi.org/10.1103/PhysRevD.84.035001}{{\em Phys. Rev. D} {\bfseries 84} (2011) 035001}, \href{https://arxiv.org/abs/1106.3101}{{\ttfamily arXiv:1106.3101 [hep-ph]}}.

\bibitem{Frigerio:2012uc}
M.~Frigerio, A.~Pomarol, F.~Riva, and A.~Urbano, ``{Composite Scalar Dark Matter},'' \href{https://dx.doi.org/10.1007/JHEP07(2012)015}{{\em JHEP} {\bfseries 07} (2012) 015}, \href{https://arxiv.org/abs/1204.2808}{{\ttfamily arXiv:1204.2808 [hep-ph]}}.

\bibitem{Buckley:2012ky}
M.~R. Buckley and E.~T. Neil, ``{Thermal dark matter from a confining sector},'' \href{https://dx.doi.org/10.1103/PhysRevD.87.043510}{{\em Phys. Rev. D} {\bfseries 87} no.~4, (2013) 043510}, \href{https://arxiv.org/abs/1209.6054}{{\ttfamily arXiv:1209.6054 [hep-ph]}}.

\bibitem{Antipin:2014qva}
O.~Antipin, M.~Redi, and A.~Strumia, ``{Dynamical generation of the weak and Dark Matter scales from strong interactions},'' \href{https://dx.doi.org/10.1007/JHEP01(2015)157}{{\em JHEP} {\bfseries 01} (2015) 157}, \href{https://arxiv.org/abs/1410.1817}{{\ttfamily arXiv:1410.1817 [hep-ph]}}.

\bibitem{Yamanaka:2014pva}
N.~Yamanaka, S.~Fujibayashi, S.~Gongyo, and H.~Iida, ``{Dark matter in the hidden gauge theory},'' \href{https://arxiv.org/abs/1411.2172}{{\ttfamily arXiv:1411.2172 [hep-ph]}}.

\bibitem{Appelquist:2015yfa}
T.~Appelquist {\em et~al.}, ``{Stealth Dark Matter: Dark scalar baryons through the Higgs portal},'' \href{https://dx.doi.org/10.1103/PhysRevD.92.075030}{{\em Phys. Rev. D} {\bfseries 92} no.~7, (2015) 075030}, \href{https://arxiv.org/abs/1503.04203}{{\ttfamily arXiv:1503.04203 [hep-ph]}}.

\bibitem{Carmona:2015haa}
A.~Carmona and M.~Chala, ``{Composite Dark Sectors},'' \href{https://dx.doi.org/10.1007/JHEP06(2015)105}{{\em JHEP} {\bfseries 06} (2015) 105}, \href{https://arxiv.org/abs/1504.00332}{{\ttfamily arXiv:1504.00332 [hep-ph]}}.

\bibitem{Soni:2016gzf}
A.~Soni and Y.~Zhang, ``{Hidden SU(N) Glueball Dark Matter},'' \href{https://dx.doi.org/10.1103/PhysRevD.93.115025}{{\em Phys. Rev. D} {\bfseries 93} no.~11, (2016) 115025}, \href{https://arxiv.org/abs/1602.00714}{{\ttfamily arXiv:1602.00714 [hep-ph]}}.

\bibitem{Harigaya:2016nlg}
K.~Harigaya, M.~Ibe, K.~Kaneta, W.~Nakano, and M.~Suzuki, ``{Thermal Relic Dark Matter Beyond the Unitarity Limit},'' \href{https://dx.doi.org/10.1007/JHEP08(2016)151}{{\em JHEP} {\bfseries 08} (2016) 151}, \href{https://arxiv.org/abs/1606.00159}{{\ttfamily arXiv:1606.00159 [hep-ph]}}.

\bibitem{Dienes:2016vei}
K.~R. Dienes, F.~Huang, S.~Su, and B.~Thomas, ``{Dynamical Dark Matter from Strongly-Coupled Dark Sectors},'' \href{https://dx.doi.org/10.1103/PhysRevD.95.043526}{{\em Phys. Rev. D} {\bfseries 95} no.~4, (2017) 043526}, \href{https://arxiv.org/abs/1610.04112}{{\ttfamily arXiv:1610.04112 [hep-ph]}}.

\bibitem{DeLuca:2018mzn}
V.~De~Luca, A.~Mitridate, M.~Redi, J.~Smirnov, and A.~Strumia, ``{Colored Dark Matter},'' \href{https://dx.doi.org/10.1103/PhysRevD.97.115024}{{\em Phys. Rev. D} {\bfseries 97} no.~11, (2018) 115024}, \href{https://arxiv.org/abs/1801.01135}{{\ttfamily arXiv:1801.01135 [hep-ph]}}.

\bibitem{Kribs:2018oad}
G.~D. Kribs, A.~Martin, and T.~Tong, ``{Effective Theories of Dark Mesons with Custodial Symmetry},'' \href{https://dx.doi.org/10.1007/JHEP08(2019)020}{{\em JHEP} {\bfseries 08} (2019) 020}, \href{https://arxiv.org/abs/1809.10183}{{\ttfamily arXiv:1809.10183 [hep-ph]}}.

\bibitem{Beylin:2019gtw}
V.~Beylin, M.~Y. Khlopov, V.~Kuksa, and N.~Volchanskiy, ``{Hadronic and Hadron-Like Physics of Dark Matter},'' \href{https://dx.doi.org/10.3390/sym11040587}{{\em Symmetry} {\bfseries 11} no.~4, (2019) 587}, \href{https://arxiv.org/abs/1904.12013}{{\ttfamily arXiv:1904.12013 [hep-ph]}}.

\bibitem{Contino:2020god}
R.~Contino, A.~Podo, and F.~Revello, ``{Composite Dark Matter from Strongly-Interacting Chiral Dynamics},'' \href{https://dx.doi.org/10.1007/JHEP02(2021)091}{{\em JHEP} {\bfseries 02} (2021) 091}, \href{https://arxiv.org/abs/2008.10607}{{\ttfamily arXiv:2008.10607 [hep-ph]}}.

\bibitem{Morrison:2020yeg}
L.~Morrison, S.~Profumo, and D.~J. Robinson, ``{Large $N$-ightmare Dark Matter},'' \href{https://dx.doi.org/10.1088/1475-7516/2021/05/058}{{\em JCAP} {\bfseries 05} (2021) 058}, \href{https://arxiv.org/abs/2010.03586}{{\ttfamily arXiv:2010.03586 [hep-ph]}}.

\bibitem{Contino:2020tix}
R.~Contino, K.~Max, and R.~K. Mishra, ``{Searching for elusive dark sectors with terrestrial and celestial observations},'' \href{https://dx.doi.org/10.1007/JHEP06(2021)127}{{\em JHEP} {\bfseries 06} (2021) 127}, \href{https://arxiv.org/abs/2012.08537}{{\ttfamily arXiv:2012.08537 [hep-ph]}}.

\bibitem{Mitridate:2017oky}
A.~Mitridate, M.~Redi, J.~Smirnov, and A.~Strumia, ``{Dark Matter as a weakly coupled Dark Baryon},'' \href{https://dx.doi.org/10.1007/JHEP10(2017)210}{{\em JHEP} {\bfseries 10} (2017) 210}, \href{https://arxiv.org/abs/1707.05380}{{\ttfamily arXiv:1707.05380 [hep-ph]}}.

\bibitem{Acharya:2017szw}
B.~S. Acharya, M.~Fairbairn, and E.~Hardy, ``{Glueball dark matter in non-standard cosmologies},'' \href{https://dx.doi.org/10.1007/JHEP07(2017)100}{{\em JHEP} {\bfseries 07} (2017) 100}, \href{https://arxiv.org/abs/1704.01804}{{\ttfamily arXiv:1704.01804 [hep-ph]}}.

\bibitem{Gross:2018zha}
C.~Gross, A.~Mitridate, M.~Redi, J.~Smirnov, and A.~Strumia, ``{Cosmological Abundance of Colored Relics},'' \href{https://dx.doi.org/10.1103/PhysRevD.99.016024}{{\em Phys. Rev. D} {\bfseries 99} no.~1, (2019) 016024}, \href{https://arxiv.org/abs/1811.08418}{{\ttfamily arXiv:1811.08418 [hep-ph]}}.

\bibitem{Contino:2018crt}
R.~Contino, A.~Mitridate, A.~Podo, and M.~Redi, ``{Gluequark Dark Matter},'' \href{https://dx.doi.org/10.1007/JHEP02(2019)187}{{\em JHEP} {\bfseries 02} (2019) 187}, \href{https://arxiv.org/abs/1811.06975}{{\ttfamily arXiv:1811.06975 [hep-ph]}}.

\bibitem{Dondi:2019olm}
N.~A. Dondi, F.~Sannino, and J.~Smirnov, ``{Thermal history of composite dark matter},'' \href{https://dx.doi.org/10.1103/PhysRevD.101.103010}{{\em Phys. Rev. D} {\bfseries 101} no.~10, (2020) 103010}, \href{https://arxiv.org/abs/1905.08810}{{\ttfamily arXiv:1905.08810 [hep-ph]}}.

\bibitem{Asadi:2021pwo}
P.~Asadi, E.~D. Kramer, E.~Kuflik, G.~W. Ridgway, T.~R. Slatyer, and J.~Smirnov, ``{Thermal squeezeout of dark matter},'' \href{https://dx.doi.org/10.1103/PhysRevD.104.095013}{{\em Phys. Rev. D} {\bfseries 104} no.~9, (2021) 095013}, \href{https://arxiv.org/abs/2103.09827}{{\ttfamily arXiv:2103.09827 [hep-ph]}}.

\bibitem{Bottaro:2021aal}
S.~Bottaro, M.~Costa, and O.~Popov, ``{Asymmetric accidental composite dark matter},'' \href{https://dx.doi.org/10.1007/JHEP11(2021)055}{{\em JHEP} {\bfseries 11} (2021) 055}, \href{https://arxiv.org/abs/2104.14244}{{\ttfamily arXiv:2104.14244 [hep-ph]}}.

\bibitem{Baldes:2021aph}
I.~Baldes, Y.~Gouttenoire, F.~Sala, and G.~Servant, ``{Supercool composite Dark Matter beyond 100 TeV},'' \href{https://dx.doi.org/10.1007/JHEP07(2022)084}{{\em JHEP} {\bfseries 07} (2022) 084}, \href{https://arxiv.org/abs/2110.13926}{{\ttfamily arXiv:2110.13926 [hep-ph]}}.

\bibitem{Asadi:2022vkc}
P.~Asadi, E.~D. Kramer, E.~Kuflik, T.~R. Slatyer, and J.~Smirnov, ``{Glueballs in a thermal squeezeout model},'' \href{https://dx.doi.org/10.1007/JHEP07(2022)006}{{\em JHEP} {\bfseries 07} (2022) 006}, \href{https://arxiv.org/abs/2203.15813}{{\ttfamily arXiv:2203.15813 [hep-ph]}}.

\bibitem{Carenza:2022pjd}
P.~Carenza, R.~Pasechnik, G.~Salinas, and Z.-W. Wang, ``{Glueball Dark Matter Revisited},'' \href{https://dx.doi.org/10.1103/PhysRevLett.129.261302}{{\em Phys. Rev. Lett.} {\bfseries 129} no.~26, (2022) 261302}, \href{https://arxiv.org/abs/2207.13716}{{\ttfamily arXiv:2207.13716 [hep-ph]}}.

\bibitem{Gouttenoire:2023roe}
Y.~Gouttenoire, E.~Kuflik, and D.~Liu, ``{Heavy baryon dark matter from SU(N) confinement: Bubble wall velocity and boundary effects},'' \href{https://dx.doi.org/10.1103/PhysRevD.109.035002}{{\em Phys. Rev. D} {\bfseries 109} no.~3, (2024) 035002}, \href{https://arxiv.org/abs/2311.00029}{{\ttfamily arXiv:2311.00029 [hep-ph]}}.

\bibitem{Strassler:2006im}
M.~J. Strassler and K.~M. Zurek, ``{Echoes of a hidden valley at hadron colliders},'' \href{https://dx.doi.org/10.1016/j.physletb.2007.06.055}{{\em Phys. Lett. B} {\bfseries 651} (2007) 374--379}, \href{https://arxiv.org/abs/hep-ph/0604261}{{\ttfamily arXiv:hep-ph/0604261}}.

\bibitem{Han:2007ae}
T.~Han, Z.~Si, K.~M. Zurek, and M.~J. Strassler, ``{Phenomenology of hidden valleys at hadron colliders},'' \href{https://dx.doi.org/10.1088/1126-6708/2008/07/008}{{\em JHEP} {\bfseries 07} (2008) 008}, \href{https://arxiv.org/abs/0712.2041}{{\ttfamily arXiv:0712.2041 [hep-ph]}}.

\bibitem{Kang:2008ea}
J.~Kang and M.~A. Luty, ``{Macroscopic Strings and 'Quirks' at Colliders},'' \href{https://dx.doi.org/10.1088/1126-6708/2009/11/065}{{\em JHEP} {\bfseries 11} (2009) 065}, \href{https://arxiv.org/abs/0805.4642}{{\ttfamily arXiv:0805.4642 [hep-ph]}}.

\bibitem{Juknevich:2009ji}
J.~E. Juknevich, D.~Melnikov, and M.~J. Strassler, ``{A Pure-Glue Hidden Valley I. States and Decays},'' \href{https://dx.doi.org/10.1088/1126-6708/2009/07/055}{{\em JHEP} {\bfseries 07} (2009) 055}, \href{https://arxiv.org/abs/0903.0883}{{\ttfamily arXiv:0903.0883 [hep-ph]}}.

\bibitem{Kilic:2009mi}
C.~Kilic, T.~Okui, and R.~Sundrum, ``{Vectorlike Confinement at the LHC},'' \href{https://dx.doi.org/10.1007/JHEP02(2010)018}{{\em JHEP} {\bfseries 02} (2010) 018}, \href{https://arxiv.org/abs/0906.0577}{{\ttfamily arXiv:0906.0577 [hep-ph]}}.

\bibitem{Juknevich:2009gg}
J.~E. Juknevich, ``{Pure-glue hidden valleys through the Higgs portal},'' \href{https://dx.doi.org/10.1007/JHEP08(2010)121}{{\em JHEP} {\bfseries 08} (2010) 121}, \href{https://arxiv.org/abs/0911.5616}{{\ttfamily arXiv:0911.5616 [hep-ph]}}.

\bibitem{Harnik:2011mv}
R.~Harnik, G.~D. Kribs, and A.~Martin, ``{Quirks at the Tevatron and Beyond},'' \href{https://dx.doi.org/10.1103/PhysRevD.84.035029}{{\em Phys. Rev. D} {\bfseries 84} (2011) 035029}, \href{https://arxiv.org/abs/1106.2569}{{\ttfamily arXiv:1106.2569 [hep-ph]}}.

\bibitem{Schwaller:2015gea}
P.~Schwaller, D.~Stolarski, and A.~Weiler, ``{Emerging Jets},'' \href{https://dx.doi.org/10.1007/JHEP05(2015)059}{{\em JHEP} {\bfseries 05} (2015) 059}, \href{https://arxiv.org/abs/1502.05409}{{\ttfamily arXiv:1502.05409 [hep-ph]}}.

\bibitem{Cohen:2015toa}
T.~Cohen, M.~Lisanti, and H.~K. Lou, ``{Semivisible Jets: Dark Matter Undercover at the LHC},'' \href{https://dx.doi.org/10.1103/PhysRevLett.115.171804}{{\em Phys. Rev. Lett.} {\bfseries 115} no.~17, (2015) 171804}, \href{https://arxiv.org/abs/1503.00009}{{\ttfamily arXiv:1503.00009 [hep-ph]}}.

\bibitem{Knapen:2016hky}
S.~Knapen, S.~Pagan~Griso, M.~Papucci, and D.~J. Robinson, ``{Triggering Soft Bombs at the LHC},'' \href{https://dx.doi.org/10.1007/JHEP08(2017)076}{{\em JHEP} {\bfseries 08} (2017) 076}, \href{https://arxiv.org/abs/1612.00850}{{\ttfamily arXiv:1612.00850 [hep-ph]}}.

\bibitem{Knapen:2017kly}
S.~Knapen, H.~K. Lou, M.~Papucci, and J.~Setford, ``{Tracking down Quirks at the Large Hadron Collider},'' \href{https://dx.doi.org/10.1103/PhysRevD.96.115015}{{\em Phys. Rev. D} {\bfseries 96} no.~11, (2017) 115015}, \href{https://arxiv.org/abs/1708.02243}{{\ttfamily arXiv:1708.02243 [hep-ph]}}.

\bibitem{Evans:2018jmd}
J.~A. Evans and M.~A. Luty, ``{Stopping Quirks at the LHC},'' \href{https://dx.doi.org/10.1007/JHEP06(2019)090}{{\em JHEP} {\bfseries 06} (2019) 090}, \href{https://arxiv.org/abs/1811.08903}{{\ttfamily arXiv:1811.08903 [hep-ph]}}.

\bibitem{Kribs:2018ilo}
G.~D. Kribs, A.~Martin, B.~Ostdiek, and T.~Tong, ``{Dark Mesons at the LHC},'' \href{https://dx.doi.org/10.1007/JHEP07(2019)133}{{\em JHEP} {\bfseries 07} (2019) 133}, \href{https://arxiv.org/abs/1809.10184}{{\ttfamily arXiv:1809.10184 [hep-ph]}}.

\bibitem{Knapen:2021eip}
S.~Knapen, J.~Shelton, and D.~Xu, ``{Perturbative benchmark models for a dark shower search program},'' \href{https://dx.doi.org/10.1103/PhysRevD.103.115013}{{\em Phys. Rev. D} {\bfseries 103} no.~11, (2021) 115013}, \href{https://arxiv.org/abs/2103.01238}{{\ttfamily arXiv:2103.01238 [hep-ph]}}.

\bibitem{Kuwahara:2023vfc}
T.~Kuwahara and S.-R. Yuan, ``{Dark vector mesons at LHC forward detector searches},'' \href{https://dx.doi.org/10.1007/JHEP06(2023)208}{{\em JHEP} {\bfseries 06} (2023) 208}, \href{https://arxiv.org/abs/2303.03736}{{\ttfamily arXiv:2303.03736 [hep-ph]}}.

\bibitem{Batz:2023zef}
A.~Batz, T.~Cohen, D.~Curtin, C.~Gemmell, and G.~D. Kribs, ``{Dark sector glueballs at the LHC},'' \href{https://dx.doi.org/10.1007/JHEP04(2024)070}{{\em JHEP} {\bfseries 04} (2024) 070}, \href{https://arxiv.org/abs/2310.13731}{{\ttfamily arXiv:2310.13731 [hep-ph]}}.

\bibitem{Bagnasco:1993st}
J.~Bagnasco, M.~Dine, and S.~D. Thomas, ``{Detecting technibaryon dark matter},'' \href{https://dx.doi.org/10.1016/0370-2693(94)90830-3}{{\em Phys. Lett. B} {\bfseries 320} (1994) 99--104}, \href{https://arxiv.org/abs/hep-ph/9310290}{{\ttfamily arXiv:hep-ph/9310290}}.

\bibitem{Alves:2009nf}
D.~S.~M. Alves, S.~R. Behbahani, P.~Schuster, and J.~G. Wacker, ``{Composite Inelastic Dark Matter},'' \href{https://dx.doi.org/10.1016/j.physletb.2010.08.006}{{\em Phys. Lett. B} {\bfseries 692} (2010) 323--326}, \href{https://arxiv.org/abs/0903.3945}{{\ttfamily arXiv:0903.3945 [hep-ph]}}.

\bibitem{SpierMoreiraAlves:2010err}
D.~Spier Moreira~Alves, S.~R. Behbahani, P.~Schuster, and J.~G. Wacker, ``{The Cosmology of Composite Inelastic Dark Matter},'' \href{https://dx.doi.org/10.1007/JHEP06(2010)113}{{\em JHEP} {\bfseries 06} (2010) 113}, \href{https://arxiv.org/abs/1003.4729}{{\ttfamily arXiv:1003.4729 [hep-ph]}}.

\bibitem{Bhattacharya:2013kma}
S.~Bhattacharya, B.~Meli\'c, and J.~Wudka, ``{Pionic Dark Matter},'' \href{https://dx.doi.org/10.1007/JHEP02(2014)115}{{\em JHEP} {\bfseries 02} (2014) 115}, \href{https://arxiv.org/abs/1307.2647}{{\ttfamily arXiv:1307.2647 [hep-ph]}}.

\bibitem{Hardy:2015boa}
E.~Hardy, R.~Lasenby, J.~March-Russell, and S.~M. West, ``{Signatures of Large Composite Dark Matter States},'' \href{https://dx.doi.org/10.1007/JHEP07(2015)133}{{\em JHEP} {\bfseries 07} (2015) 133}, \href{https://arxiv.org/abs/1504.05419}{{\ttfamily arXiv:1504.05419 [hep-ph]}}.

\bibitem{Detmold:2014qqa}
W.~Detmold, M.~McCullough, and A.~Pochinsky, ``{Dark Nuclei I: Cosmology and Indirect Detection},'' \href{https://dx.doi.org/10.1103/PhysRevD.90.115013}{{\em Phys. Rev. D} {\bfseries 90} no.~11, (2014) 115013}, \href{https://arxiv.org/abs/1406.2276}{{\ttfamily arXiv:1406.2276 [hep-ph]}}.

\bibitem{Mahbubani:2019pij}
R.~Mahbubani, M.~Redi, and A.~Tesi, ``{Indirect detection of composite asymmetric dark matter},'' \href{https://dx.doi.org/10.1103/PhysRevD.101.103037}{{\em Phys. Rev. D} {\bfseries 101} no.~10, (2020) 103037}, \href{https://arxiv.org/abs/1908.00538}{{\ttfamily arXiv:1908.00538 [hep-ph]}}.

\bibitem{Cline:2013zca}
J.~M. Cline, Z.~Liu, G.~D. Moore, and W.~Xue, ``{Composite strongly interacting dark matter},'' \href{https://dx.doi.org/10.1103/PhysRevD.90.015023}{{\em Phys. Rev. D} {\bfseries 90} no.~1, (2014) 015023}, \href{https://arxiv.org/abs/1312.3325}{{\ttfamily arXiv:1312.3325 [hep-ph]}}.

\bibitem{Boddy:2014yra}
K.~K. Boddy, J.~L. Feng, M.~Kaplinghat, and T.~M.~P. Tait, ``{Self-Interacting Dark Matter from a Non-Abelian Hidden Sector},'' \href{https://dx.doi.org/10.1103/PhysRevD.89.115017}{{\em Phys. Rev. D} {\bfseries 89} no.~11, (2014) 115017}, \href{https://arxiv.org/abs/1402.3629}{{\ttfamily arXiv:1402.3629 [hep-ph]}}.

\bibitem{Krnjaic:2014xza}
G.~Krnjaic and K.~Sigurdson, ``{Big Bang Darkleosynthesis},'' \href{https://dx.doi.org/10.1016/j.physletb.2015.11.001}{{\em Phys. Lett. B} {\bfseries 751} (2015) 464--468}, \href{https://arxiv.org/abs/1406.1171}{{\ttfamily arXiv:1406.1171 [hep-ph]}}.

\bibitem{Buen-Abad:2015ova}
M.~A. Buen-Abad, G.~Marques-Tavares, and M.~Schmaltz, ``{Non-Abelian dark matter and dark radiation},'' \href{https://dx.doi.org/10.1103/PhysRevD.92.023531}{{\em Phys. Rev. D} {\bfseries 92} no.~2, (2015) 023531}, \href{https://arxiv.org/abs/1505.03542}{{\ttfamily arXiv:1505.03542 [hep-ph]}}.

\bibitem{Garani:2021zrr}
R.~Garani, M.~Redi, and A.~Tesi, ``{Dark QCD matters},'' \href{https://dx.doi.org/10.1007/JHEP12(2021)139}{{\em JHEP} {\bfseries 12} (2021) 139}, \href{https://arxiv.org/abs/2105.03429}{{\ttfamily arXiv:2105.03429 [hep-ph]}}.

\bibitem{LatticeStrongDynamicsLSD:2013elk}
{\bfseries Lattice Strong Dynamics (LSD)} Collaboration, T.~Appelquist {\em et~al.}, ``{Lattice Calculation of Composite Dark Matter Form Factors},'' \href{https://dx.doi.org/10.1103/PhysRevD.88.014502}{{\em Phys. Rev. D} {\bfseries 88} no.~1, (2013) 014502}, \href{https://arxiv.org/abs/1301.1693}{{\ttfamily arXiv:1301.1693 [hep-ph]}}.

\bibitem{LSD:2014obp}
{\bfseries LSD} Collaboration, T.~Appelquist {\em et~al.}, ``{Composite bosonic baryon dark matter on the lattice: SU(4) baryon spectrum and the effective Higgs interaction},'' \href{https://dx.doi.org/10.1103/PhysRevD.89.094508}{{\em Phys. Rev. D} {\bfseries 89} no.~9, (2014) 094508}, \href{https://arxiv.org/abs/1402.6656}{{\ttfamily arXiv:1402.6656 [hep-lat]}}.

\bibitem{Detmold:2014kba}
W.~Detmold, M.~McCullough, and A.~Pochinsky, ``{Dark nuclei. II. Nuclear spectroscopy in two-color QCD},'' \href{https://dx.doi.org/10.1103/PhysRevD.90.114506}{{\em Phys. Rev. D} {\bfseries 90} no.~11, (2014) 114506}, \href{https://arxiv.org/abs/1406.4116}{{\ttfamily arXiv:1406.4116 [hep-lat]}}.

\bibitem{Appelquist:2015zfa}
T.~Appelquist {\em et~al.}, ``{Detecting Stealth Dark Matter Directly through Electromagnetic Polarizability},'' \href{https://dx.doi.org/10.1103/PhysRevLett.115.171803}{{\em Phys. Rev. Lett.} {\bfseries 115} no.~17, (2015) 171803}, \href{https://arxiv.org/abs/1503.04205}{{\ttfamily arXiv:1503.04205 [hep-ph]}}.

\bibitem{Francis:2018xjd}
A.~Francis, R.~J. Hudspith, R.~Lewis, and S.~Tulin, ``{Dark Matter from Strong Dynamics: The Minimal Theory of Dark Baryons},'' \href{https://dx.doi.org/10.1007/JHEP12(2018)118}{{\em JHEP} {\bfseries 12} (2018) 118}, \href{https://arxiv.org/abs/1809.09117}{{\ttfamily arXiv:1809.09117 [hep-ph]}}.

\bibitem{LatticeStrongDynamics:2020jwi}
{\bfseries Lattice Strong Dynamics} Collaboration, R.~C. Brower, K.~Cushman, {\em et~al.}, ``{Stealth dark matter confinement transition and gravitational waves},'' \href{https://dx.doi.org/10.1103/PhysRevD.103.014505}{{\em Phys. Rev. D} {\bfseries 103} no.~1, (2021) 014505}, \href{https://arxiv.org/abs/2006.16429}{{\ttfamily arXiv:2006.16429 [hep-lat]}}.

\bibitem{Brower:2023rqf}
R.~C. Brower, C.~Culver, {\em et~al.}, ``{Stealth dark matter spectrum using LapH and Irreps},'' \href{https://arxiv.org/abs/2312.07836}{{\ttfamily arXiv:2312.07836 [hep-lat]}}.

\bibitem{Kribs:2016cew}
G.~D. Kribs and E.~T. Neil, ``{Review of strongly-coupled composite dark matter models and lattice simulations},'' \href{https://dx.doi.org/10.1142/S0217751X16430041}{{\em Int. J. Mod. Phys. A} {\bfseries 31} no.~22, (2016) 1643004}, \href{https://arxiv.org/abs/1604.04627}{{\ttfamily arXiv:1604.04627 [hep-ph]}}.

\bibitem{Cacciapaglia:2020kgq}
G.~Cacciapaglia, C.~Pica, and F.~Sannino, ``{Fundamental Composite Dynamics: A Review},'' \href{https://dx.doi.org/10.1016/j.physrep.2020.07.002}{{\em Phys. Rept.} {\bfseries 877} (2020) 1--70}, \href{https://arxiv.org/abs/2002.04914}{{\ttfamily arXiv:2002.04914 [hep-ph]}}.

\bibitem{Cline:2021itd}
J.~M. Cline, ``{Dark atoms and composite dark matter},'' \href{https://dx.doi.org/10.21468/SciPostPhysLectNotes.52}{{\em SciPost Phys. Lect. Notes} {\bfseries 52} (2022) 1}, \href{https://arxiv.org/abs/2108.10314}{{\ttfamily arXiv:2108.10314 [hep-ph]}}.

\bibitem{Asadi:2022njl}
P.~Asadi {\em et~al.}, ``{Early-Universe Model Building},'' \href{https://arxiv.org/abs/2203.06680}{{\ttfamily arXiv:2203.06680 [hep-ph]}}.

\bibitem{LZ:2022lsv}
{\bfseries LZ} Collaboration, J.~Aalbers {\em et~al.}, ``{First Dark Matter Search Results from the LUX-ZEPLIN (LZ) Experiment},'' \href{https://dx.doi.org/10.1103/PhysRevLett.131.041002}{{\em Phys. Rev. Lett.} {\bfseries 131} no.~4, (2023) 041002}, \href{https://arxiv.org/abs/2207.03764}{{\ttfamily arXiv:2207.03764 [hep-ex]}}.

\bibitem{LZ2024}
{\bfseries LZ Collaboration} Collaboration, S.~Haselschwardt, ``{\href{https://indico.uchicago.edu/event/427/contributions/1325/attachments/359/548/lz_results_tevpa.pdf}{New Dark Matter Search Results from the LUX-ZEPLIN (LZ) Experiment}},'' {\em Talk Presented at TeVPA 2024} .

\bibitem{Krall:2017xij}
R.~Krall and M.~Reece, ``{Last Electroweak WIMP Standing: Pseudo-Dirac Higgsino Status and Compact Stars as Future Probes},'' \href{https://dx.doi.org/10.1088/1674-1137/42/4/043105}{{\em Chin. Phys. C} {\bfseries 42} no.~4, (2018) 043105}, \href{https://arxiv.org/abs/1705.04843}{{\ttfamily arXiv:1705.04843 [hep-ph]}}.

\bibitem{Kavanagh:2018xeh}
B.~J. Kavanagh, P.~Panci, and R.~Ziegler, ``{Faint Light from Dark Matter: Classifying and Constraining Dark Matter-Photon Effective Operators},'' \href{https://dx.doi.org/10.1007/JHEP04(2019)089}{{\em JHEP} {\bfseries 04} (2019) 089}, \href{https://arxiv.org/abs/1810.00033}{{\ttfamily arXiv:1810.00033 [hep-ph]}}.

\bibitem{Essig:2007az}
R.~Essig, ``{Direct Detection of Non-Chiral Dark Matter},'' \href{https://dx.doi.org/10.1103/PhysRevD.78.015004}{{\em Phys. Rev. D} {\bfseries 78} (2008) 015004}, \href{https://arxiv.org/abs/0710.1668}{{\ttfamily arXiv:0710.1668 [hep-ph]}}.

\bibitem{Hisano:2010fy}
J.~Hisano, K.~Ishiwata, and N.~Nagata, ``{A complete calculation for direct detection of Wino dark matter},'' \href{https://dx.doi.org/10.1016/j.physletb.2010.05.047}{{\em Phys. Lett. B} {\bfseries 690} (2010) 311--315}, \href{https://arxiv.org/abs/1004.4090}{{\ttfamily arXiv:1004.4090 [hep-ph]}}.

\bibitem{Hill:2014yka}
R.~J. Hill and M.~P. Solon, ``{Standard Model anatomy of WIMP dark matter direct detection I: weak-scale matching},'' \href{https://dx.doi.org/10.1103/PhysRevD.91.043504}{{\em Phys. Rev. D} {\bfseries 91} (2015) 043504}, \href{https://arxiv.org/abs/1401.3339}{{\ttfamily arXiv:1401.3339 [hep-ph]}}.

\bibitem{Hill:2014yxa}
R.~J. Hill and M.~P. Solon, ``{Standard Model anatomy of WIMP dark matter direct detection II: QCD analysis and hadronic matrix elements},'' \href{https://dx.doi.org/10.1103/PhysRevD.91.043505}{{\em Phys. Rev. D} {\bfseries 91} (2015) 043505}, \href{https://arxiv.org/abs/1409.8290}{{\ttfamily arXiv:1409.8290 [hep-ph]}}.

\bibitem{Chen:2018uqz}
C.-Y. Chen, R.~J. Hill, M.~P. Solon, and A.~M. Wijangco, ``{Power Corrections to the Universal Heavy WIMP-Nucleon Cross Section},'' \href{https://dx.doi.org/10.1016/j.physletb.2018.04.021}{{\em Phys. Lett. B} {\bfseries 781} (2018) 473--479}, \href{https://arxiv.org/abs/1801.08551}{{\ttfamily arXiv:1801.08551 [hep-ph]}}.

\bibitem{Chen:2023bwg}
Q.~Chen, G.-J. Ding, and R.~J. Hill, ``{General heavy WIMP nucleon elastic scattering},'' \href{https://dx.doi.org/10.1103/PhysRevD.108.116023}{{\em Phys. Rev. D} {\bfseries 108} no.~11, (2023) 116023}, \href{https://arxiv.org/abs/2309.02715}{{\ttfamily arXiv:2309.02715 [hep-ph]}}.

\bibitem{Pospelov:2000bq}
M.~Pospelov and T.~ter Veldhuis, ``{Direct and indirect limits on the electromagnetic form-factors of WIMPs},'' \href{https://dx.doi.org/10.1016/S0370-2693(00)00358-0}{{\em Phys. Lett. B} {\bfseries 480} (2000) 181--186}, \href{https://arxiv.org/abs/hep-ph/0003010}{{\ttfamily arXiv:hep-ph/0003010}}.

\bibitem{Sigurdson:2004zp}
K.~Sigurdson, M.~Doran, A.~Kurylov, R.~R. Caldwell, and M.~Kamionkowski, ``{Dark-matter electric and magnetic dipole moments},'' \href{https://dx.doi.org/10.1103/PhysRevD.70.083501}{{\em Phys. Rev. D} {\bfseries 70} (2004) 083501}, \href{https://arxiv.org/abs/astro-ph/0406355}{{\ttfamily arXiv:astro-ph/0406355}}. [Erratum: Phys.Rev.D 73, 089903 (2006)].

\bibitem{Masso:2009mu}
E.~Masso, S.~Mohanty, and S.~Rao, ``{Dipolar Dark Matter},'' \href{https://dx.doi.org/10.1103/PhysRevD.80.036009}{{\em Phys. Rev. D} {\bfseries 80} (2009) 036009}, \href{https://arxiv.org/abs/0906.1979}{{\ttfamily arXiv:0906.1979 [hep-ph]}}.

\bibitem{Barger:2010gv}
V.~Barger, W.-Y. Keung, and D.~Marfatia, ``{Electromagnetic properties of dark matter: Dipole moments and charge form factor},'' \href{https://dx.doi.org/10.1016/j.physletb.2010.12.008}{{\em Phys. Lett. B} {\bfseries 696} (2011) 74--78}, \href{https://arxiv.org/abs/1007.4345}{{\ttfamily arXiv:1007.4345 [hep-ph]}}.

\bibitem{Banks:2010eh}
T.~Banks, J.-F. Fortin, and S.~Thomas, ``{Direct Detection of Dark Matter Electromagnetic Dipole Moments},'' \href{https://arxiv.org/abs/1007.5515}{{\ttfamily arXiv:1007.5515 [hep-ph]}}.

\bibitem{DelNobile:2012tx}
E.~Del~Nobile, C.~Kouvaris, P.~Panci, F.~Sannino, and J.~Virkajarvi, ``{Light Magnetic Dark Matter in Direct Detection Searches},'' \href{https://dx.doi.org/10.1088/1475-7516/2012/08/010}{{\em JCAP} {\bfseries 08} (2012) 010}, \href{https://arxiv.org/abs/1203.6652}{{\ttfamily arXiv:1203.6652 [hep-ph]}}.

\bibitem{Weiner:2012cb}
N.~Weiner and I.~Yavin, ``{How Dark Are Majorana WIMPs? Signals from MiDM and Rayleigh Dark Matter},'' \href{https://dx.doi.org/10.1103/PhysRevD.86.075021}{{\em Phys. Rev. D} {\bfseries 86} (2012) 075021}, \href{https://arxiv.org/abs/1206.2910}{{\ttfamily arXiv:1206.2910 [hep-ph]}}.

\bibitem{Weiner:2012gm}
N.~Weiner and I.~Yavin, ``{UV completions of magnetic inelastic and Rayleigh dark matter for the Fermi Line(s)},'' \href{https://dx.doi.org/10.1103/PhysRevD.87.023523}{{\em Phys. Rev. D} {\bfseries 87} no.~2, (2013) 023523}, \href{https://arxiv.org/abs/1209.1093}{{\ttfamily arXiv:1209.1093 [hep-ph]}}.

\bibitem{DelNobile:2014eta}
E.~Del~Nobile, G.~B. Gelmini, P.~Gondolo, and J.-H. Huh, ``{Direct detection of Light Anapole and Magnetic Dipole DM},'' \href{https://dx.doi.org/10.1088/1475-7516/2014/06/002}{{\em JCAP} {\bfseries 06} (2014) 002}, \href{https://arxiv.org/abs/1401.4508}{{\ttfamily arXiv:1401.4508 [hep-ph]}}.

\bibitem{Arina:2020mxo}
C.~Arina, A.~Cheek, K.~Mimasu, and L.~Pagani, ``{Light and Darkness: consistently coupling dark matter to photons via effective operators},'' \href{https://dx.doi.org/10.1140/epjc/s10052-021-09010-1}{{\em Eur. Phys. J. C} {\bfseries 81} no.~3, (2021) 223}, \href{https://arxiv.org/abs/2005.12789}{{\ttfamily arXiv:2005.12789 [hep-ph]}}.

\bibitem{DelNobile:2021wmp}
E.~Del~Nobile, ``{The Theory of Direct Dark Matter Detection: A Guide to Computations},'' \href{https://arxiv.org/abs/2104.12785}{{\ttfamily arXiv:2104.12785 [hep-ph]}}.

\bibitem{Hambye:2021xvd}
T.~Hambye and X.-J. Xu, ``{Dark matter electromagnetic dipoles: the WIMP expectation},'' \href{https://dx.doi.org/10.1007/JHEP11(2021)156}{{\em JHEP} {\bfseries 11} (2021) 156}, \href{https://arxiv.org/abs/2106.01403}{{\ttfamily arXiv:2106.01403 [hep-ph]}}.

\bibitem{PICO:2022ohk}
{\bfseries PICO} Collaboration, B.~Ali {\em et~al.}, ``{Results on photon-mediated dark-matter\textendash{}nucleus interactions from the PICO-60 C3F8 bubble chamber},'' \href{https://dx.doi.org/10.1103/PhysRevD.106.042004}{{\em Phys. Rev. D} {\bfseries 106} no.~4, (2022) 042004}, \href{https://arxiv.org/abs/2204.10340}{{\ttfamily arXiv:2204.10340 [astro-ph.CO]}}.

\bibitem{PandaX:2023toi}
{\bfseries PandaX} Collaboration, X.~Ning {\em et~al.}, ``{Limits on the luminance of dark matter from xenon recoil data},'' \href{https://dx.doi.org/10.1038/s41586-023-05982-0}{{\em Nature} {\bfseries 618} no.~7963, (2023) 47--50}.

\bibitem{Eby:2023wem}
J.~Eby, P.~J. Fox, and G.~D. Kribs, ``{Earth-catalyzed detection of magnetic inelastic dark matter with photons in large underground detectors},'' \href{https://dx.doi.org/10.1007/JHEP06(2024)165}{{\em JHEP} {\bfseries 06} (2024) 165}, \href{https://arxiv.org/abs/2312.08478}{{\ttfamily arXiv:2312.08478 [hep-ph]}}.

\bibitem{Griest:1989wd}
K.~Griest and M.~Kamionkowski, ``{Unitarity Limits on the Mass and Radius of Dark Matter Particles},'' \href{https://dx.doi.org/10.1103/PhysRevLett.64.615}{{\em Phys. Rev. Lett.} {\bfseries 64} (1990) 615}.

\bibitem{vonHarling:2014kha}
B.~von Harling and K.~Petraki, ``{Bound-state formation for thermal relic dark matter and unitarity},'' \href{https://dx.doi.org/10.1088/1475-7516/2014/12/033}{{\em JCAP} {\bfseries 12} (2014) 033}, \href{https://arxiv.org/abs/1407.7874}{{\ttfamily arXiv:1407.7874 [hep-ph]}}.

\bibitem{Smirnov:2019ngs}
J.~Smirnov and J.~F. Beacom, ``{TeV-Scale Thermal WIMPs: Unitarity and its Consequences},'' \href{https://dx.doi.org/10.1103/PhysRevD.100.043029}{{\em Phys. Rev. D} {\bfseries 100} no.~4, (2019) 043029}, \href{https://arxiv.org/abs/1904.11503}{{\ttfamily arXiv:1904.11503 [hep-ph]}}.

\bibitem{Georgi:1984zwz}
H.~Georgi, {\em {Weak Interactions and Modern Particle Theory}}.
\newblock 1984.

\bibitem{Abe:2024mwa}
T.~Abe, R.~Sato, and T.~Yamanaka, ``{Composite Dark Matter with Forbidden Annihilation},'' \href{https://arxiv.org/abs/2404.03963}{{\ttfamily arXiv:2404.03963 [hep-ph]}}.

\bibitem{ABK}
P.~Asadi, A.~Batz, and G.~Kribs, ``{in preparation},''.

\bibitem{Vafa:1983tf}
C.~Vafa and E.~Witten, ``{Restrictions on Symmetry Breaking in Vector-Like Gauge Theories},'' \href{https://dx.doi.org/10.1016/0550-3213(84)90230-X}{{\em Nucl. Phys. B} {\bfseries 234} (1984) 173--188}.

\bibitem{Ovanesyan:2014fha}
G.~Ovanesyan and L.~Vecchi, ``{Direct detection of dark matter polarizability},'' \href{https://dx.doi.org/10.1007/JHEP07(2015)128}{{\em JHEP} {\bfseries 07} (2015) 128}, \href{https://arxiv.org/abs/1410.0601}{{\ttfamily arXiv:1410.0601 [hep-ph]}}.

\bibitem{Berlin:2018tvf}
A.~Berlin, N.~Blinov, S.~Gori, P.~Schuster, and N.~Toro, ``{Cosmology and Accelerator Tests of Strongly Interacting Dark Matter},'' \href{https://dx.doi.org/10.1103/PhysRevD.97.055033}{{\em Phys. Rev. D} {\bfseries 97} no.~5, (2018) 055033}, \href{https://arxiv.org/abs/1801.05805}{{\ttfamily arXiv:1801.05805 [hep-ph]}}.

\bibitem{Bernreuther:2019pfb}
E.~Bernreuther, F.~Kahlhoefer, M.~Kr\"amer, and P.~Tunney, ``{Strongly interacting dark sectors in the early Universe and at the LHC through a simplified portal},'' \href{https://dx.doi.org/10.1007/JHEP01(2020)162}{{\em JHEP} {\bfseries 01} (2020) 162}, \href{https://arxiv.org/abs/1907.04346}{{\ttfamily arXiv:1907.04346 [hep-ph]}}.

\bibitem{Frandsen:2012db}
M.~T. Frandsen, U.~Haisch, F.~Kahlhoefer, P.~Mertsch, and K.~Schmidt-Hoberg, ``{Loop-induced dark matter direct detection signals from gamma-ray lines},'' \href{https://dx.doi.org/10.1088/1475-7516/2012/10/033}{{\em JCAP} {\bfseries 10} (2012) 033}, \href{https://arxiv.org/abs/1207.3971}{{\ttfamily arXiv:1207.3971 [hep-ph]}}.

\bibitem{Bottaro:2021snn}
S.~Bottaro, D.~Buttazzo, M.~Costa, R.~Franceschini, P.~Panci, D.~Redigolo, and L.~Vittorio, ``{Closing the window on WIMP Dark Matter},'' \href{https://dx.doi.org/10.1140/epjc/s10052-021-09917-9}{{\em Eur. Phys. J. C} {\bfseries 82} no.~1, (2022) 31}, \href{https://arxiv.org/abs/2107.09688}{{\ttfamily arXiv:2107.09688 [hep-ph]}}.

\bibitem{Bottaro:2022one}
S.~Bottaro, D.~Buttazzo, M.~Costa, R.~Franceschini, P.~Panci, D.~Redigolo, and L.~Vittorio, ``{The last complex WIMPs standing},'' \href{https://dx.doi.org/10.1140/epjc/s10052-022-10918-5}{{\em Eur. Phys. J. C} {\bfseries 82} no.~11, (2022) 992}, \href{https://arxiv.org/abs/2205.04486}{{\ttfamily arXiv:2205.04486 [hep-ph]}}.

\bibitem{Bloch:2024suj}
I.~M. Bloch, S.~Bottaro, D.~Redigolo, and L.~Vittorio, ``{Looking for WIMPs through the neutrino fogs},'' \href{https://arxiv.org/abs/2410.02723}{{\ttfamily arXiv:2410.02723 [hep-ph]}}.

\bibitem{Mitridate:2017izz}
A.~Mitridate, M.~Redi, J.~Smirnov, and A.~Strumia, ``{Cosmological Implications of Dark Matter Bound States},'' \href{https://dx.doi.org/10.1088/1475-7516/2017/05/006}{{\em JCAP} {\bfseries 05} (2017) 006}, \href{https://arxiv.org/abs/1702.01141}{{\ttfamily arXiv:1702.01141 [hep-ph]}}.

\bibitem{DeRujula:1975qlm}
A.~De~Rujula, H.~Georgi, and S.~L. Glashow, ``{Hadron Masses in a Gauge Theory},'' \href{https://dx.doi.org/10.1103/PhysRevD.12.147}{{\em Phys. Rev. D} {\bfseries 12} (1975) 147--162}.

\bibitem{Manohar:1983md}
A.~Manohar and H.~Georgi, ``{Chiral Quarks and the Nonrelativistic Quark Model},'' \href{https://dx.doi.org/10.1016/0550-3213(84)90231-1}{{\em Nucl. Phys. B} {\bfseries 234} (1984) 189--212}.

\bibitem{Workman:2022ynf}
{\bfseries Particle Data Group} Collaboration, R.~L. Workman and Others, ``{Review of Particle Physics},'' \href{https://dx.doi.org/10.1093/ptep/ptac097}{{\em PTEP} {\bfseries 2022} (2022) 083C01}.

\bibitem{Aranda:2015jis}
A.~Aranda, L.~Barajas, and J.~A.~R. Cembranos, ``{Magnetic dipole moments for composite dark matter},'' \href{https://dx.doi.org/10.1088/1475-7516/2016/03/034}{{\em JCAP} {\bfseries 03} (2016) 034}, \href{https://arxiv.org/abs/1511.02805}{{\ttfamily arXiv:1511.02805 [hep-ph]}}.

\bibitem{Antipin:2015xia}
O.~Antipin, M.~Redi, A.~Strumia, and E.~Vigiani, ``{Accidental Composite Dark Matter},'' \href{https://dx.doi.org/10.1007/JHEP07(2015)039}{{\em JHEP} {\bfseries 07} (2015) 039}, \href{https://arxiv.org/abs/1503.08749}{{\ttfamily arXiv:1503.08749 [hep-ph]}}.

\bibitem{Tucker-Smith:2001myb}
D.~Tucker-Smith and N.~Weiner, ``{Inelastic dark matter},'' \href{https://dx.doi.org/10.1103/PhysRevD.64.043502}{{\em Phys. Rev. D} {\bfseries 64} (2001) 043502}, \href{https://arxiv.org/abs/hep-ph/0101138}{{\ttfamily arXiv:hep-ph/0101138}}.

\bibitem{Chang:2008gd}
S.~Chang, G.~D. Kribs, D.~Tucker-Smith, and N.~Weiner, ``{Inelastic Dark Matter in Light of DAMA/LIBRA},'' \href{https://dx.doi.org/10.1103/PhysRevD.79.043513}{{\em Phys. Rev. D} {\bfseries 79} (2009) 043513}, \href{https://arxiv.org/abs/0807.2250}{{\ttfamily arXiv:0807.2250 [hep-ph]}}.

\bibitem{Chang:2010en}
S.~Chang, N.~Weiner, and I.~Yavin, ``{Magnetic Inelastic Dark Matter},'' \href{https://dx.doi.org/10.1103/PhysRevD.82.125011}{{\em Phys. Rev. D} {\bfseries 82} (2010) 125011}, \href{https://arxiv.org/abs/1007.4200}{{\ttfamily arXiv:1007.4200 [hep-ph]}}.

\bibitem{Kumar:2011iy}
K.~Kumar, A.~Menon, and T.~M.~P. Tait, ``{Magnetic Fluffy Dark Matter},'' \href{https://dx.doi.org/10.1007/JHEP02(2012)131}{{\em JHEP} {\bfseries 02} (2012) 131}, \href{https://arxiv.org/abs/1111.2336}{{\ttfamily arXiv:1111.2336 [hep-ph]}}.

\bibitem{Bramante:2016rdh}
J.~Bramante, P.~J. Fox, G.~D. Kribs, and A.~Martin, ``{Inelastic frontier: Discovering dark matter at high recoil energy},'' \href{https://dx.doi.org/10.1103/PhysRevD.94.115026}{{\em Phys. Rev.} {\bfseries D94} no.~11, (2016) 115026},
\href{https://arxiv.org/abs/1608.02662}{{\ttfamily arXiv:1608.02662 [hep-ph]}}.

\bibitem{Feldstein:2010su}
B.~Feldstein, P.~W. Graham, and S.~Rajendran, ``{Luminous Dark Matter},'' \href{https://dx.doi.org/10.1103/PhysRevD.82.075019}{{\em Phys. Rev. D} {\bfseries 82} (2010) 075019}, \href{https://arxiv.org/abs/1008.1988}{{\ttfamily arXiv:1008.1988 [hep-ph]}}.

\bibitem{Eby:2019mgs}
J.~Eby, P.~J. Fox, R.~Harnik, and G.~D. Kribs, ``{Luminous Signals of Inelastic Dark Matter in Large Detectors},'' \href{https://dx.doi.org/10.1007/JHEP09(2019)115}{{\em JHEP} {\bfseries 09} (2019) 115}, \href{https://arxiv.org/abs/1904.09994}{{\ttfamily arXiv:1904.09994 [hep-ph]}}.

\bibitem{Graham:2024syw}
P.~W. Graham, H.~Ramani, and S.~S.~Y. Wong, ``{Enhancing Direct Detection of Higgsino Dark Matter},'' \href{https://arxiv.org/abs/2409.07768}{{\ttfamily arXiv:2409.07768 [hep-ph]}}.

\bibitem{PhysRevD.40.2997}
G.~Morpurgo, ``Field theory and the nonrelativistic quark model: A parametrization of the baryon magnetic moments and masses,'' \href{https://dx.doi.org/10.1103/PhysRevD.40.2997}{{\em Phys. Rev. D} {\bfseries 40} (Nov, 1989) 2997--3011}. \url{https://link.aps.org/doi/10.1103/PhysRevD.40.2997}.

\bibitem{PhysRevD.41.2865}
G.~Morpurgo, ``Field theory and the nonrelativistic quark model: A parametrization of the meson masses,'' \href{https://dx.doi.org/10.1103/PhysRevD.41.2865}{{\em Phys. Rev. D} {\bfseries 41} (May, 1990) 2865--2870}. \url{https://link.aps.org/doi/10.1103/PhysRevD.41.2865}.

\bibitem{PhysRevD.53.3754}
G.~Dillon and G.~Morpurgo, ``Relation of constituent quark models to qcd: Why several simple models work ``so well'','' \href{https://dx.doi.org/10.1103/PhysRevD.53.3754}{{\em Phys. Rev. D} {\bfseries 53} (Apr, 1996) 3754--3769}. \url{https://link.aps.org/doi/10.1103/PhysRevD.53.3754}.

\bibitem{Buchmann:2000wf}
A.~J. Buchmann and R.~F. Lebed, ``{Large N(c), constituent quarks, and N, Delta charge radii},'' \href{https://dx.doi.org/10.1103/PhysRevD.62.096005}{{\em Phys. Rev. D} {\bfseries 62} (2000) 096005}, \href{https://arxiv.org/abs/hep-ph/0003167}{{\ttfamily arXiv:hep-ph/0003167}}.

\bibitem{Cheng:1998hc}
H.-C. Cheng, B.~A. Dobrescu, and K.~T. Matchev, ``{Generic and chiral extensions of the supersymmetric standard model},'' \href{https://dx.doi.org/10.1016/S0550-3213(99)00012-7}{{\em Nucl. Phys. B} {\bfseries 543} (1999) 47--72}, \href{https://arxiv.org/abs/hep-ph/9811316}{{\ttfamily arXiv:hep-ph/9811316}}.

\bibitem{Gherghetta:1999sw}
T.~Gherghetta, G.~F. Giudice, and J.~D. Wells, ``{Phenomenological consequences of supersymmetry with anomaly induced masses},'' \href{https://dx.doi.org/10.1016/S0550-3213(99)00429-0}{{\em Nucl. Phys. B} {\bfseries 559} (1999) 27--47}, \href{https://arxiv.org/abs/hep-ph/9904378}{{\ttfamily arXiv:hep-ph/9904378}}.

\bibitem{Feng:1999fu}
J.~L. Feng, T.~Moroi, L.~Randall, M.~Strassler, and S.-f. Su, ``{Discovering supersymmetry at the Tevatron in wino LSP scenarios},'' \href{https://dx.doi.org/10.1103/PhysRevLett.83.1731}{{\em Phys. Rev. Lett.} {\bfseries 83} (1999) 1731--1734}, \href{https://arxiv.org/abs/hep-ph/9904250}{{\ttfamily arXiv:hep-ph/9904250}}.

\bibitem{Cirelli:2005uq}
M.~Cirelli, N.~Fornengo, and A.~Strumia, ``{Minimal dark matter},'' \href{https://dx.doi.org/10.1016/j.nuclphysb.2006.07.012}{{\em Nucl. Phys. B} {\bfseries 753} (2006) 178--194}, \href{https://arxiv.org/abs/hep-ph/0512090}{{\ttfamily arXiv:hep-ph/0512090}}.

\bibitem{Gasser:1982ap}
J.~Gasser and H.~Leutwyler, ``{Quark Masses},'' \href{https://dx.doi.org/10.1016/0370-1573(82)90035-7}{{\em Phys. Rept.} {\bfseries 87} (1982) 77--169}.

\bibitem{Jedamzik:2006xz}
K.~Jedamzik, ``{Big bang nucleosynthesis constraints on hadronically and electromagnetically decaying relic neutral particles},'' \href{https://dx.doi.org/10.1103/PhysRevD.74.103509}{{\em Phys. Rev. D} {\bfseries 74} (2006) 103509}, \href{https://arxiv.org/abs/hep-ph/0604251}{{\ttfamily arXiv:hep-ph/0604251}}.

\bibitem{Kawasaki:2017bqm}
M.~Kawasaki, K.~Kohri, T.~Moroi, and Y.~Takaesu, ``{Revisiting Big-Bang Nucleosynthesis Constraints on Long-Lived Decaying Particles},'' \href{https://dx.doi.org/10.1103/PhysRevD.97.023502}{{\em Phys. Rev. D} {\bfseries 97} no.~2, (2018) 023502}, \href{https://arxiv.org/abs/1709.01211}{{\ttfamily arXiv:1709.01211 [hep-ph]}}.

\bibitem{Morningstar:1999rf}
C.~J. Morningstar and M.~J. Peardon, ``{The Glueball spectrum from an anisotropic lattice study},'' \href{https://dx.doi.org/10.1103/PhysRevD.60.034509}{{\em Phys. Rev. D} {\bfseries 60} (1999) 034509}, \href{https://arxiv.org/abs/hep-lat/9901004}{{\ttfamily arXiv:hep-lat/9901004}}.

\bibitem{Mathieu:2008me}
V.~Mathieu, N.~Kochelev, and V.~Vento, ``{The Physics of Glueballs},'' \href{https://dx.doi.org/10.1142/S0218301309012124}{{\em Int. J. Mod. Phys. E} {\bfseries 18} (2009) 1--49}, \href{https://arxiv.org/abs/0810.4453}{{\ttfamily arXiv:0810.4453 [hep-ph]}}.

\bibitem{Cheng:2021kjg}
H.-C. Cheng, L.~Li, and E.~Salvioni, ``{A theory of dark pions},'' \href{https://dx.doi.org/10.1007/JHEP01(2022)122}{{\em JHEP} {\bfseries 01} (2022) 122}, \href{https://arxiv.org/abs/2110.10691}{{\ttfamily arXiv:2110.10691 [hep-ph]}}.

\bibitem{Dienes:2021cxr}
K.~R. Dienes, D.~Kim, T.~T. Leininger, and B.~Thomas, ``{Sequential displaced vertices: Novel collider signature for long-lived particles},'' \href{https://dx.doi.org/10.1103/PhysRevD.106.095012}{{\em Phys. Rev. D} {\bfseries 106} no.~9, (2022) 095012}, \href{https://arxiv.org/abs/2108.02204}{{\ttfamily arXiv:2108.02204 [hep-ph]}}.

\bibitem{Cohen:2013ama}
T.~Cohen, M.~Lisanti, A.~Pierce, and T.~R. Slatyer, ``{Wino Dark Matter Under Siege},'' \href{https://dx.doi.org/10.1088/1475-7516/2013/10/061}{{\em JCAP} {\bfseries 10} (2013) 061}, \href{https://arxiv.org/abs/1307.4082}{{\ttfamily arXiv:1307.4082 [hep-ph]}}.

\bibitem{Asadi:2016ybp}
P.~Asadi, M.~Baumgart, P.~J. Fitzpatrick, E.~Krupczak, and T.~R. Slatyer, ``{Capture and Decay of Electroweak WIMPonium},'' \href{https://dx.doi.org/10.1088/1475-7516/2017/02/005}{{\em JCAP} {\bfseries 02} (2017) 005}, \href{https://arxiv.org/abs/1610.07617}{{\ttfamily arXiv:1610.07617 [hep-ph]}}.

\bibitem{Baumgart:2017nsr}
M.~Baumgart, T.~Cohen, I.~Moult, N.~L. Rodd, T.~R. Slatyer, M.~P. Solon, I.~W. Stewart, and V.~Vaidya, ``{Resummed Photon Spectra for WIMP Annihilation},'' \href{https://dx.doi.org/10.1007/JHEP03(2018)117}{{\em JHEP} {\bfseries 03} (2018) 117}, \href{https://arxiv.org/abs/1712.07656}{{\ttfamily arXiv:1712.07656 [hep-ph]}}.

\bibitem{Baumgart:2018yed}
M.~Baumgart, T.~Cohen, E.~Moulin, I.~Moult, L.~Rinchiuso, N.~L. Rodd, T.~R. Slatyer, I.~W. Stewart, and V.~Vaidya, ``{Precision Photon Spectra for Wino Annihilation},'' \href{https://dx.doi.org/10.1007/JHEP01(2019)036}{{\em JHEP} {\bfseries 01} (2019) 036}, \href{https://arxiv.org/abs/1808.08956}{{\ttfamily arXiv:1808.08956 [hep-ph]}}.

\bibitem{Baumgart:2023pwn}
M.~Baumgart, N.~L. Rodd, T.~R. Slatyer, and V.~Vaidya, ``{The quintuplet annihilation spectrum},'' \href{https://dx.doi.org/10.1007/JHEP01(2024)158}{{\em JHEP} {\bfseries 01} (2024) 158}, \href{https://arxiv.org/abs/2309.11562}{{\ttfamily arXiv:2309.11562 [hep-ph]}}.

\bibitem{Leane:2021ihh}
R.~K. Leane, T.~Linden, P.~Mukhopadhyay, and N.~Toro, ``{Celestial-Body Focused Dark Matter Annihilation Throughout the Galaxy},'' \href{https://dx.doi.org/10.1103/PhysRevD.103.075030}{{\em Phys. Rev. D} {\bfseries 103} no.~7, (2021) 075030}, \href{https://arxiv.org/abs/2101.12213}{{\ttfamily arXiv:2101.12213 [astro-ph.HE]}}.

\bibitem{Parreno:2016fwu}
A.~Parreno, M.~J. Savage, B.~C. Tiburzi, J.~Wilhelm, E.~Chang, W.~Detmold, and K.~Orginos, ``{Octet baryon magnetic moments from lattice QCD: Approaching experiment from a three-flavor symmetric point},'' \href{https://dx.doi.org/10.1103/PhysRevD.95.114513}{{\em Phys. Rev. D} {\bfseries 95} no.~11, (2017) 114513}, \href{https://arxiv.org/abs/1609.03985}{{\ttfamily arXiv:1609.03985 [hep-lat]}}.

\end{thebibliography}\endgroup
\bibliographystyle{utphys}

\end{document}